\documentclass[runningheads,11points]{llncs}
\usepackage[utf8, latin5]{inputenc}
\usepackage{amsmath,amssymb}
\usepackage{color}
\usepackage{cite}
\usepackage{fancyhdr}
\usepackage{graphicx}
\usepackage{colortbl}
\usepackage{dsfont,hyperref}
\usepackage{vmargin}
 \usepackage{subfigure}
\setmarginsrb{4.0cm}{1.5cm}{4.0cm}{3.5cm}{1.2cm}{1.3cm}{0.9cm}{1.2cm}

\usepackage[lined,boxed,commentsnumbered]{algorithm2e}

\spnewtheorem{observation}{Observation}{\bfseries}{\itshape}
\spnewtheorem{claimN}{Claim}{\bfseries}{\itshape}
\spnewtheorem{remarkLI}{Remark}{\bfseries}{\itshape}
\renewenvironment{proof}{\noindent \textbf{Proof:}}{\hfill$\square$\medskip}
\newenvironment{proofof}{\noindent \textbf{Proof:}}{\hfill$\Diamond$\medskip}

\newcommand{\tw}{\mbox{\bf tw}}
\newcommand{\tcw}{\mbox{\bf tcw}}
\newcommand{\td}{\mbox{\bf td}}
\newcommand{\cx}{\mathcal{X}}
\newcommand{\dd}{\mathcal{D}}

\newcommand{\tcd}{(T,\cx)}
\newcommand{\fpt}{{\sf FPT}}
\newcommand{\xp}{{\sf XP}}
\newcommand{\prbcc}{\textsc{MinCCA}}
\newcommand{\rc} {\textit{rc}}
\newcommand{\cc}{\textit{cc}}
\newcommand{\TT}{{\cal T}}

\renewcommand{\root}[1]{{\sf root}(#1)}

\newcommand{\cut}{\mbox{\sf cut}}
\newcommand{\adh}{\mbox{\sf adh}}
\newcommand{\tor}{\mbox{\sf tor}}

\newcommand{\supprOK}[1]{}

\newcommand{\remove}[1]{}

\newcommand{\commentfig}[1]{#1}
\usepackage{tikz}

\newcommand{\probl}[3]{
\begin{flushleft}
\fbox{
\begin{minipage}{\linewidth}
\noindent {\sc #1}\\
          {\bf Input:} #2\\
          {\bf Output:} #3
\end{minipage}}
\medskip
\end{flushleft}
}
\newcommand{\paraprobl}[4]
{
  \begin{flushleft}
    \fbox{
      \begin{minipage}{.96\linewidth}
        \noindent {\textsc {#1}}\\
        {\bf Input:} #2\\
        {\bf Parameter:} #4\\
        {\bf Question:} #3
      \end{minipage}
    }
  \end{flushleft}
}

\usepackage{amsmath}
\usepackage{amssymb}

\newcommand {\npc}     {\textsc{NP}\textrm{-complete}}
\newcommand {\nph}     {\textsc{NP}\textrm{-hard}}

\newcommand {\abs}[1]{\left\vert#1\right\vert}
\newcommand {\set}[1]{\left\{#1\right\}}

\newcommand {\defined} {:=}
\title{Parameterized complexity of the MINCCA problem  on graphs of bounded decomposability\thanks{An extended abstract of this article will appear in the \emph{Proceedings of the 42nd International Workshop on Graph-Theoretic Concepts in Computer Science (\textbf{WG}), Istanbul, Turkey, June \textbf{2016}}. This work is supported by the bilateral research program of CNRS and TUBITAK under grant no.114E731.}}

\author{Didem G\"{o}z\"{u}pek~\inst{1}, Sibel \"{O}zkan~\inst{2}, Christophe Paul~\inst{3},\\Ignasi Sau~\inst{3}, and Mordechai Shalom~\inst{4,5}\thanks{The work of this author is supported in part by the TUBITAK 2221 Programme.}
}

\authorrunning{D. G\"{o}z\"{u}pek, S. \"{O}zkan, C. Paul, I. Sau, and M. Shalom}
\titlerunning{Parameterized complexity of the MINCCA problem}

\institute{Department of Computer Engineering, Gebze Technical University, Kocaeli, Turkey\\
\email{didem.gozupek@gtu.edu.tr}
\and
Department of Mathematics, Gebze Technical University, Kocaeli, Turkey\\
\email{s.ozkan@gtu.edu.tr}
\and
CNRS, LIRMM, Universit\'e de Montpellier, Montpellier, France\\
\email{paul@lirmm.fr, sau@lirmm.fr}
\and
TelHai College, Upper Galilee, 12210, Israel\\
\email{cmshalom@telhai.ac.il}
\and
Department of Industrial Engineering, Bo\u{g}azi\c{c}i University, Istanbul, Turkey
}

\begin{document}

\maketitle
\begin{abstract}
In an edge-colored graph, the cost incurred at a vertex on a path when two incident edges with different colors are traversed is called reload or changeover cost. The \textit{Minimum Changeover Cost Arborescence} ($\prbcc$) problem consists in finding an arborescence with a given root vertex such that the total changeover cost of the internal vertices is minimized. It has been recently proved by G{\"{o}}z{\"{u}}pek \emph{et al}.~\cite{FPT-by-tw-Delta} that the $\prbcc$ problem is $\fpt$ when parameterized by the treewidth and the maximum degree of the input graph. In this article we present the following results for $\prbcc$:
\begin{itemize}
\item[$\bullet$] the problem is {\sc{W[1]}}-hard parameterized by the treedepth of the input graph, even on graphs of average degree at most 8. In particular, it is {\sc{W[1]}}-hard parameterized by the treewidth of the input graph, which answers the main open problem of~\cite{FPT-by-tw-Delta};
\item[$\bullet$] it is {\sc{W[1]}}-hard on multigraphs parameterized by the tree-cutwidth of the input multigraph;

\item[$\bullet$] it is $\fpt$ parameterized by the star tree-cutwidth of the input graph, which is a slightly restricted version of tree-cutwidth. This result strictly generalizes the $\fpt$ result given in~\cite{FPT-by-tw-Delta};

\item[$\bullet$] it remains {\sc{NP}}-hard on planar graphs even when restricted to instances with at most 6 colors and 0/1 symmetric costs, or when restricted to instances with at most 8 colors, maximum degree bounded by $4$, and 0/1 symmetric costs.

\end{itemize}

 \vspace{0.19cm}

\end{abstract}

\section{Introduction}\label{sec:intro}
The cost that occurs at a vertex when two incident edges with different colors are crossed over is referred to as \emph{reload cost} or \emph{changeover cost} in the literature. This cost depends on the colors of the traversed edges. Although the reload cost concept has important applications in numerous areas such as transportation networks, energy distribution networks, and cognitive radio networks, it has received little attention in the literature. In particular, reload/changeover cost problems have been investigated very little from the perspective of parameterized complexity; the only previous work we are aware of is the one in \cite{FPT-by-tw-Delta}.

In heterogeneous networks in telecommunications, transiting from a technology such as 3G (third generation) to another technology such as wireless local area network (WLAN) has an overhead in terms of delay, power consumption etc., depending on the particular setting. This cost has gained increasing importance due to the recently popular concept of vertical handover \cite{CGM13}, which is a technique that allows a mobile user to stay connected to the Internet (without a connection loss) by switching to a different wireless network when necessary. Likewise, switching between different service providers even if they have the same technology has a non-negligible cost. Recently, cognitive radio networks (CRN) have gained increasing attention in the communication networks research community. Unlike other wireless technologies, CRNs are envisioned to operate in a wide range of frequencies. Therefore, switching from one frequency band to another frequency band in a CRN has a significant cost in terms of delay and power consumption \cite{GBA13, AAK+13}. This concept has applications in other areas as well. For instance, the cost of transferring cargo from one mode of transportation to another has a significant cost that outweighs even the cost of transporting the cargo from one place to another using a single mode of transportation \cite{WiSt01}. In energy distribution networks, transferring energy from one type of carrier to another has an important cost corresponding to reload costs \cite{Ga08}.

The reload cost concept was introduced in \cite{WiSt01}, where the considered problem is to find a spanning tree having minimum diameter with respect to reload cost. In particular, they proved that the problem cannot be approximated within a factor better than $3$ even on graphs with maximum degree $5$, in addition to providing a polynomial-time algorithm for graphs with maximum degree $3$. The work in \cite{Ga08} extended these inapproximability results by proving that the problem is inapproximable within a factor better than $2$ even on graphs with maximum degree $4$. When reload costs satisfy the triangle inequality, they showed that the problem is inapproximable within any factor better than $5/3$.

The work in \cite{GGM14} focused on the minimum reload cost cycle cover problem, which is to find a set of vertex-disjoint cycles spanning all vertices with minimum total reload cost. They showed an inapproximability result for the case when there are $2$ colors, the reload costs are symmetric and satisfy the triangle inequality. They also presented some integer programming formulations and computational results.

The authors in \cite{GLMM10} study the problems of finding a path, trail or walk connecting two given vertices with minimum total reload cost. They present several polynomial and NP-hard cases for (a)symmetric reload costs and reload costs with(out) triangle inequality. Furthermore, they show that the problem is polynomial for walks, as previously mentioned by \cite{WiSt01}, and re-proved later for directed graphs by \cite{AGM11}.

The work in \cite{GGM11} introduced the \textit{Minimum Changeover Cost Arborescence} ($\prbcc$) problem. Given a root vertex, $\prbcc$ problem is to find an arborescence with minimum total changeover cost starting from the root vertex. They proved that even on graphs with bounded degree and reload costs adhering to the triangle inequality, $\prbcc$ on directed graphs is inapproximable within  $\beta \log\log(n)$ for $\beta>0$ when there are two colors, and within $n^{1/3-\epsilon}$ for any $\epsilon>0$ when there are three colors. The work in \cite{GozupekSVZ14} investigated several special cases of the problem such as bounded cost values, bounded degree, and bounded number of colors. In addition, \cite{GozupekSVZ14} presented inapproximability results as well as a polynomial-time algorithm and an approximation algorithm for the considered special cases.

In this paper, we study the $\prbcc$ problem from the perspective of parameterized complexity; see \cite{FG06,Nie06,DF13,CyganFKLMPPS15}. Unlike the classical complexity theory, parameterized complexity theory takes into account not only the total input size $n$, but also other aspects of the problem encoded in a parameter $k$. It mainly aims to find an exact resolution of $\npc$ problems. A problem is called \textit{fixed parameter tractable} ($\fpt$) if it can be solved in time $f(k) \cdot p(n)$, where $f(k)$ is a function depending solely on $k$ and $p(n)$ is a polynomial in $n$. An algorithm constituting such a solution is called an $\fpt$ algorithm for the problem. Analogously to NP-completeness in classical complexity, the theory of {\sc{W[1]}}-hardness can be used to show that a problem is unlikely to be $\fpt$, i.e., for every algorithm the parameter has to appear in the exponent of $n$. The parameterized complexity of reload cost problems is largely unexplored in the literature. To the best of our knowledge, \cite{FPT-by-tw-Delta} is the only work that focuses on this issue by studying the $\prbcc$ problem on bounded treewidth graphs. In particular, \cite{FPT-by-tw-Delta} showed that the $\prbcc$ problem is in $\xp$ when parameterized by the treewidth of the input graph and it is $\fpt$ when parameterized by the treewidth {\sl and} the maximum degree of the input graph. In this paper, we prove that the $\prbcc$ problem is {\sc{W[1]}}-hard parameterized by the treedepth of the input graph, even on graphs of average degree at most 8. In particular, it is {\sc{W[1]}}-hard parameterized by the treewidth of the input graph, which answers the main open issue pointed out by~\cite{FPT-by-tw-Delta}. Furthermore, we prove that it is {\sc{W[1]}}-hard on multigraphs parameterized by the tree-cutwidth of the input multigraph. On the positive side, we present an $\fpt$ algorithm parameterized by the star tree-cutwidth of the input graph, which is a slightly restricted version of tree-cutwidth that we introduce here. This algorithm strictly generalizes the $\fpt$ algorithm given in \cite{FPT-by-tw-Delta}. We also prove that the problem is $\nph$ on planar graphs, which are also graphs of bounded decomposability, even when restricted to instances with at most 6 colors and 0/1 symmetric costs. In addition, we prove that it remains $\nph$ on planar graphs even when restricted to instances with at most 8 colors, maximum degree bounded by 4, and 0/1 symmetric costs.

The rest of this paper is organized as follows. In Section \ref{sec:prelim} we introduce some basic definitions and preliminaries, as well as a formal definition of the $\prbcc$ problem. We present our hardness results in Section \ref{sec:hardness}, and  our algorithmic results with respect to star tree-cutwidth are given in Section~\ref{sec:algorithms}. Finally, Section \ref{sec:conclusion} concludes the paper.

%Due to space limitations, the proofs of the results marked with `$[\star]$', our algorithmic results with respect to star tree-cutwidth, as well as several figures, can be found in the full version of the article, which is permanently available at [\texttt{arXiv:XXXX/YYYY}].  \ig{UPDATE THIS!!}

%our algorithmic results with respect to star tree-cutwidth have been completely moved to Appendix~\ref{sec:algorithms}. Finally, Section \ref{sec:conclusion} concludes the paper. The proofs of the results marked with `$[\star]$' have been moved to the appendices. All figures except Fig.~\ref{fig:treedepthTCW-Kim} can be found in the full version of the article. % Appendix~\ref{ap:figures}.

\section{Preliminaries}\label{sec:prelim}
We say that two partial functions $f$ and $f'$ \emph{agree} if they have the same value everywhere they are both defined, and we denote it by $f \sim f'$. For a set $A$ and an element $x$, we use $A + x$ (resp., $A - x$) as a shorthand for $A \cup \set{x}$ (resp., $A \setminus \set{x}$). We denote by $[i,k]$ the set of all integers between $i$ and $k$ inclusive, and $[k]=[1,k]$.

\paragraph{\textbf{\emph{Graphs, digraphs, trees, and forests}}.} Given an undirected (multi)graph $G$ and a subset $U \subseteq V(G)$ of the vertices of $G$, $\delta_G(U) \defined \set{\{u,u'\} \in E(G) \mid u \in U, u' \notin U}$ is the \emph{cut} of $G$ determined by $U$, i.e., the set of edges of $G$ that have exactly one end in $U$. In particular, $\delta_G(v)$ denotes the set of edges incident to $v$ in $G$, and $d_G(v) \defined \abs{\delta_G(v)}$ is the \emph{degree} of $v$ in $G$. The \emph{minimum and maximum degrees} of $G$ are defined as $\delta(G)\defined \min \set{d_G(v) \mid v \in V(G)}$ and $\Delta(G)\defined \max \set{d_G(v) \mid v \in V(G)}$ respectively. We denote by $N_G(U)$ (resp., $N_G[U]$) the \emph{open} (resp., \emph{closed}) \emph{neighborhood} of $U$ in $G$. $N_G(U)$ is the set of vertices of $V(G) \setminus U$ that are adjacent to a vertex of $U$, and $N_G[U] \defined N_G(U) \cup U$. When there is no ambiguity about the graph $G$ we omit it from the subscripts. For a subset of vertices $U \subseteq V(G)$, $G[U]$ denotes the subgraph of $G$ \emph{induced} by $U$.

A digraph $T$ is a \emph{rooted tree} or \emph{arborescence} if its underlying graph is a tree and it contains a \emph{root} vertex denoted by $\root{T}$ with a directed path from every other vertex to it. Every other vertex $v \neq \root{T}$ has a parent in $T$, and $v$ is a \emph{child} of its parent.

A rooted forest is the disjoint union of rooted trees, that is, each connected component of it has a root, which will be called a \emph{sink} of the forest.

\paragraph{\textbf{\emph{Tree decompositions, treewidth, and treedepth}}.} A \emph{tree decomposition} of a graph $G=(V(G), E(G))$ is a tree $\TT$, where $V(\TT) = \set{B_1, B_2, \ldots}$ is a set of subsets (called \emph{bags}) of $V(G)$ such that the following three conditions are met:
\begin{enumerate}
    \item $\bigcup V(\TT) = V(G)$.
    \item For every edge $uv \in E(G)$, $u,v \in B_i$ for some bag $B_i \in V(\TT)$.
    \item For every $B_i, B_j, B_k \in V(\TT)$ such that $B_k$ is on the path $P_\TT(B_i, B_j)$, $B_i \cap B_j \subseteq B_k$.
\end{enumerate}

The \emph{width} $\omega(\TT)$ of a tree decomposition $\TT$ is defined as the size of its largest bag minus 1, i.e., $\omega(\TT) = \max \set{\abs{B}~|~B \in V(\TT)} - 1$. The \emph{treewidth}
of a graph $G$, denoted as $\tw(G)$, is defined as the minimum width among all tree decompositions of $G$. When the treewidth of the input graph is bounded, many efficient algorithms are known for problems that are in general $\nph$. In fact, most problems are known to be $\fpt$ when parameterized by the treewidth of the input graph. Hence, what we prove in this paper, i.e., the $\prbcc$ problem is {\sc{W[1]}}-hard when parameterized by treewidth, is an interesting result.

The \emph{treedepth} $\td (G)$ of a graph $G$ is the smallest natural number $k$ such that each vertex of $G$ can be labeled with an element from $\set{1,\dots, k}$ so that every path in $G$ joining two vertices with the same label contains a vertex having a larger label. Intuitively, where the treewidth parameter measures how far a graph is from being a tree, treedepth measures how far a graph is from being a star. The treewidth of a graph is at most one less than its treedepth; therefore, a {\sc{W[1]}}-hardness result for treedepth implies a {\sc{W[1]}}-hardness for treewidth.

%\comment{Treewidth, treedepth, and their relation}

\paragraph{\textbf{\emph{Tree-cutwidth}}.} We now explain the concept of tree-cutwidth and follow the notation in~\cite{Ganian0S15}. A \emph{tree-cut decomposition} of a graph $G$ is a pair $\tcd$ where $T$ is a rooted tree and $\cx$ is a near-partition of $V(G)$ (that is, empty sets are allowed) where each set $X_t$ of the partition is associated with a node $t$ of $T$. That is, $\cx=\set{X_t \subseteq V(G): t \in V(T)}$. The set $X_t$ is termed the \emph{bag} associated with the \emph{node} $t$. For a node $t$ of $T$ we denote by $Y_t$ the union of all the bags associated with $t$ and its descendants, and $G_t=G[Y_t]$. $\cut(t)=\delta(Y_t)$ is the set of all edges with exactly one endpoint in $Y_t$.

The \emph{adhesion} $\adh(t)$ of $t$ is $\abs{\cut(t)}$. The \emph{torso} of $t$ is the graph $H_t$ obtained from $G$ as follows. Let $t_1,\ldots,t_\ell$ be the children of $t$, $Y_i=Y_{t_i}$ for $i \in [\ell]$ and $Y_0=V(G) \setminus (X_t \cup_{i=1}^\ell Y_i)$. We first contract each set $Y_i$ to a single vertex $y_i$, by possibly creating parallel edges. We then remove every vertex $y_i$ of degree $1$ (with its incident edge), and finally \emph{suppress} every vertex $y_i$ of degree $2$ having 2 neighbors, by connecting its two neighbors with an edge and removing $y_i$.
%Given a graph $G$ and $X \subseteq V (G)$, let the \emph{$3$-center of $(G,X )$} be the unique graph obtained from $G$ by exhaustively suppressing vertices in $V ( G ) \ X$ of degree
%at most two
The \emph{torso size} $\tor(t)$ of $t$ is the number of vertices in $H_t$. The \emph{width} of a tree-cut decomposition $\tcd$  of $G$ is
$\max_{t \in V (T)}\{\adh(t),\tor (t)\}$. The\emph{ tree-cutwidth} of $G$, or $ \tcw(G)$ in short, is the
minimum width of $\tcd$ over all tree-cut decompositions $\tcd$ of $G$.

%\begin{theorem}[Kim \emph{et al}.~\cite{KOPST15}]\label{thm:KimEtAl2Approx}
% Given a graph $G$ on $n$ vertices, a tree-cut decomposition of $G$ of width at most $2 \tcw(G)$ can be computed in time $2^{O(k^2 \log {\scriptsize\tcw}(G))} \cdot n^2$.
%\end{theorem}
%
%A non-root node $t$ of $T$ is \emph{thin} if $\adh(t) \leq 2$ and \emph{bold} otherwise. A tree-cut decomposition $\tcd$ is \emph{nice} if for every thin node $t$ and every sibling $t'$ of $t$, $Y_t$ does not have neighbours in $Y_{t'}$. For a node $t$ of $T$ we use $B_t$ to denote the set of thin children $t'$  of $t$ such that $N(Y_{t'}) \subseteq X_t$, and we let $A_t$ contain every child of $t$ which is not in $B_t$.
%
%\begin{theorem}[Ganian \emph{et al}.~\cite{Ganian0S15}]\label{thm:GanianAlgorithmNiceTCWD}
% A tree-cut decomposition of a graph $G$ can be transformed in time $O(n^3)$ to a nice tree-cut decomposition of $G$ without increasing its width or its number of nodes.
%\end{theorem}
%
%\begin{lemma}[Ganian \emph{et al}.~\cite{Ganian0S15}]\label{thm:GanianAtBounded}
%For every node $t$ of a tree-cut decomposition of width $k$,
%\begin{equation}\label{eq:At}
%|A_t| \leq 2k+1.
%\end{equation}
%\end{lemma}

\begin{figure}[t]
\begin{center}
\commentfig{
%\vspace{-2cm}
%\hspace*{1.0cm}
\includegraphics[width=3.8cm]{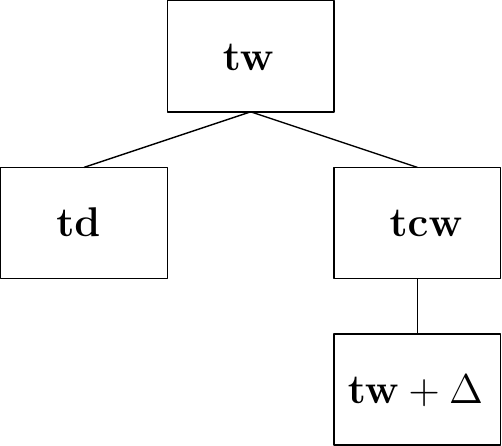}
%\vspace{-3.5cm}
}
\caption{Relationships between the graph parameters under consideration. $A$ being a child of $B$ means that every graph class with bounded $A$ has also bounded $B$, but the converse is not necessarily true \cite{Ganian0S15}.} \label{fig:treedepthTCW-Kim}
\end{center}
\end{figure}

Fig.~\ref{fig:treedepthTCW-Kim} shows the relationship between the graph parameters that we consider in this article. As depicted in Fig.~\ref{fig:treedepthTCW-Kim}, tree-cutwidth  provides an intermediate measurement which allows either to push the boundary of fixed parameter tractability or strengthen {\sc{W[1]}}-hardness result (cf.~\cite{Wollan15,Ganian0S15,KOPST15}). Furthermore, Fig.~\ref{fig:treedepthTCW-Kim} also shows that treedepth and tree-cutwidth do not possess an implication relation in terms of parameterized complexity.

\paragraph{\textbf{\emph{Reload and changeover costs}}.} We follow the notation and terminology of \cite{WiSt01} where the concept of reload cost was defined. We consider edge colored graphs $G$, where the colors are taken from a finite set $X$ and $\chi:E(G) \rightarrow X$ is the \emph{coloring function}. Given a coloring function $\chi$, we denote by $E^\chi_x$, or simply by $E_x$  the set of edges of $E$ colored $x$, and $G_x=(V(G),E(G)_x)$ is the subgraph of $G$ having the same vertex set as $G$, but only the edges colored $x$. The costs are given by a non-negative function $\cc:X^2 \rightarrow \mathbb{N}_0$ satisfying
\begin{enumerate}
\item $\cc(x_1,x_2)=\cc(x_2,x_1)$ for every $x_1,x_2 \in X$.
\item $\cc(x,x)=0$ for every $x \in X$.
\end{enumerate}
%At this point we note that our algorithms does not rely on any one of the above properties of the cost function, and they will work on other cost functions too.
%We assume without loss of generality that the minimum non-zero cost value is $1$ and we denote the maximum cost value as $C_{max} \defined \max_{x_1, x_2 \in X} \cc(x_1,x_2)$.
The cost of traversing two incident edges $e_1,e_2$ is $\cc(e_1,e_2) \defined \cc(\chi(e_1),\chi(e_2))$.

We say that an instance satisfies the {\it triangle inequality}, if (in addition to the above) the cost function satisfies
$\cc(e_1,e_3) \leq \cc(e_1,e_2)+\cc(e_2,e_3)$ whenever $e_1,e_2$ and $e_3$ are incident to the same vertex.

The changeover cost of a path $P=(e_1 - e_2 - \ldots - e_\ell)$ of length $\ell$ is $\cc(P) \defined \sum_{i=2}^{\ell} \cc(e_{i-1},e_i)$. Note that $\cc(P)=0$ whenever $\ell \leq 1$.

We extend this definition to trees as follows: Given a directed tree $T$ rooted at $r$, (resp., an undirected tree $T$ and a vertex $r \in V(T)$), for every outgoing edge $e$ of $r$ (resp., incident to $r$) we define $prev(e)=e$, and for every other edge $prev(e)$ is the edge preceding $e$ on the path from $r$ to $e$. The changeover cost of $T$ with respect to $r$ is $\cc(T, r)\defined\sum_{e \in E(T)} \cc(prev(e),e)$. When there is no ambiguity about the vertex $r$, we denote $\cc(T,r)$ by $\cc(T)$.

\newpage

\paragraph{\textbf{\emph{Statement of the problem}}.} The $\prbcc$ problem aims to find a spanning tree rooted at $r$ with minimum changeover cost\cite{GGM11}. Formally,

\probl
{$\prbcc$}
{A graph $G=(V,E)$ with an edge coloring function $\chi:E \rightarrow X$, a vertex $r \in V$ and a changeover cost function $\cc:X^2 \rightarrow \mathbb{N}_0$.}
{A spanning tree $T$ of $G$ minimizing $\cc(T,r)$.}

\section{Hardness results}\label{sec:hardness}
In this section we prove several hardness results for the $\prbcc$ problem. Our main result is in Subsection~\ref{sec:hard-bounded-average-degree}, where we prove that the problem is {\sc{W[1]}}-hard parameterized by the treedepth of the input graph. We also prove that the problem is {\sc{W[1]}}-hard on {\sl multigraphs} parameterized by the tree-cutwidth of the input graph. Both results hold even if the input graph has bounded average degree.  Finally, in Subsection~\ref{sec:hard-planar} we prove that the problem remains {\sf NP}-hard on planar graphs.

\subsection{W[1]-hardness with parameters treedepth and tree-cutwidth}\label{sec:hard-bounded-average-degree}
We need to define the following parameterized problem.

\paraprobl
{\textsc{Multicolored $k$-Clique}}
{A graph $G$, a coloring function $c: V(G) \to \{1,\ldots,k\}$, and a positive integer $k$.}
{Does $G$ contain a clique on $k$ vertices with one vertex from each color class?}
{$k$.}

\vspace{.2cm}

\textsc{Multicolored $k$-Clique} is known to be {\sc{W[1]}}-hard on general graphs, even in the special case where all color classes have the same number of
vertices~\cite{Pietrzak03}, and therefore we may make this assumption as well.

\begin{theorem}\label{thm:hardness-treewidth}
The $\prbcc$ problem is {\sc{W[1]}}-hard parameterized by the treedepth of the input graph, even on graphs with average degree at most 8.
%\ig{even on graphs of average degree at most 8}
\end{theorem}
\begin{proof} We reduce from \textsc{Multicolored $k$-Clique}, where we may assume that $k$ is odd. Indeed, given an instance $(G,c,k)$ of \textsc{Multicolored $k$-Clique}, we can trivially reduce the problem to itself as follows. If $k$ is odd, we do nothing. Otherwise, we output $(G',c',k+1)$, where $G'$ is obtained from $G$ by adding a universal vertex $v$, and $c':V(G') \to \{1,\ldots,k+1\}$ is such that its restriction to $G$ equals $c$, and $c(v) = k+1$.

Given an instance $(G,c,k)$ of \textsc{Multicolored $k$-Clique} with $k$ odd, we proceed to construct an instance $(H,X,\chi,r,\cc)$ of \prbcc. Let $V(G) = V_1 \uplus V_2 \uplus \dots \uplus V_k$, where the vertices of $V_i$ are colored $i$ for $1 \leq i \leq k$. Let $W$ be an arbitrary Eulerian circuit of the complete graph $K_k$, which exists since $k$ is odd. If $V(K_k)=\{v_1,\ldots,v_k\}$, we can clearly assume without loss of generality\footnote{This assumption is not crucial for the construction, but helps in making it conceptually and notationally easier.} that $W$ starts by visiting, in this order, vertices $v_1,v_2,\ldots,v_k,v_1$, and that the last edge of $W$ is $\{v_3,v_1\}$; see Fig.~\ref{fig:main-a} and Fig.~\ref{fig:main-b}. For every edge $\{v_i,v_j\}$ of $W$, we add to $H$ a vertex $s_{i,j}$. These vertices are called the \emph{selector vertices} of $H$. For every two consecutive edges $\{v_i,v_j\}, \{v_j,v_{\ell}\}$ of $W$, we add to $H$ a vertex $v_j^{i,\ell}$ and we make it adjacent to both $s_{i,j}$ and $s_{j,\ell}$. We also add to $H$ a new vertex $v_1^{0,2}$ adjacent to $s_{1,2}$, a new vertex $v_1^{3,0}$ adjacent to $s_{3,1}$, and a new vertex $r$ adjacent to $v_1^{0,2}$, which will be the \emph{root} of $H$. Note that the graph constructed so far is a simple path $P$ on $2{k \choose 2} + 2$  vertices; see Fig.~\ref{fig:main-c}. We say that the vertices of the form $v_j^{i,\ell}$ are \emph{occurrences} of vertex $v_j \in V(K_k)$.  For $2 \leq j \leq k$, we add an edge between the root $r$ and the first occurrence of vertex $v_j$ in $P$ (note that the edge between $r$ and the first occurrence of $v_1$ already exists).

\vspace{-.6cm}
\begin{figure}[h!]
  \centering
  \includegraphics[scale=.12]{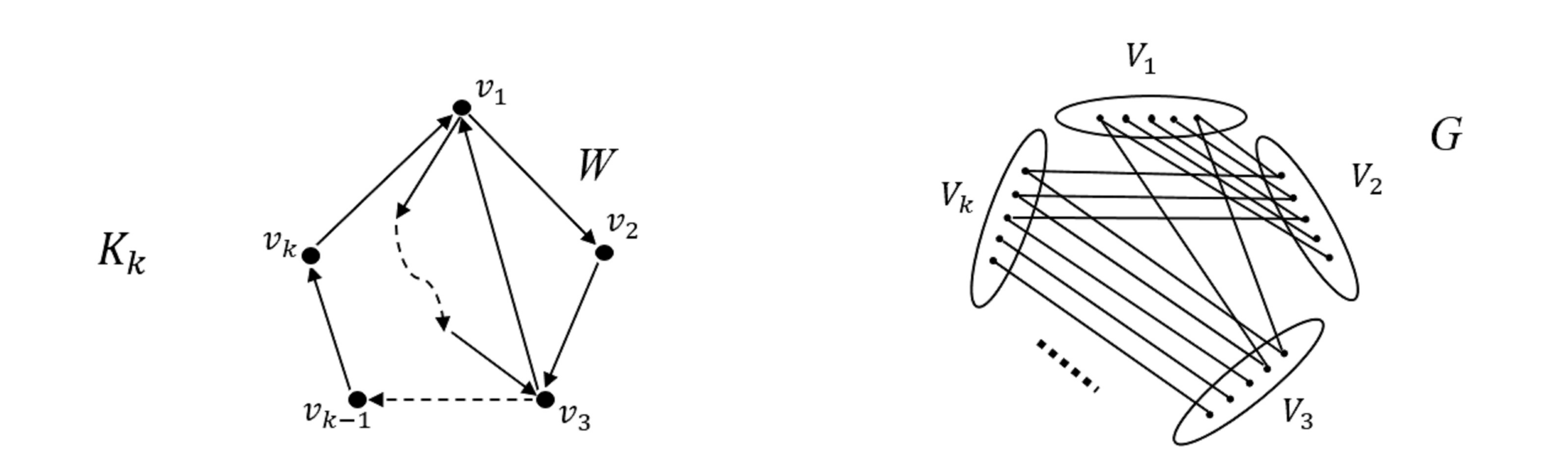}
  \caption{The complete graph $K_k$ and an Eulerian circuit $W$ in $K_k$ starting with $v_1,v_2,\ldots,v_k,v_1$ and ending with $v_3,v_1$. A $k$-colored graph $G$ is also illustrated.} \label{fig:main-a}
\end{figure}

%\vspace{.4cm}

\begin{figure}[h!]
  \centering
  \hspace*{-2.5cm}\includegraphics[scale=0.1]{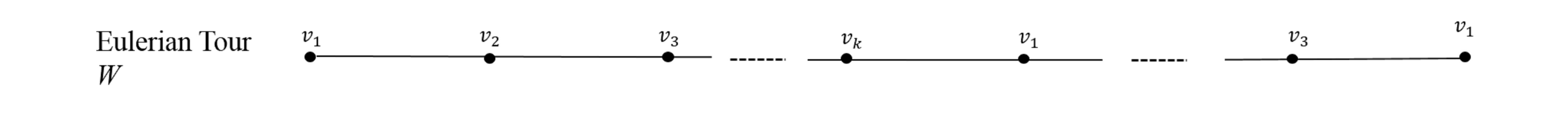}
  \caption{A left-to-right representation of the Eulerian circuit $W$.} \label{fig:main-b}
\end{figure}

%\vspace{.2cm}

\begin{figure}[h!]
  \centering
 \hspace*{-2.2cm}\includegraphics[scale=0.12]{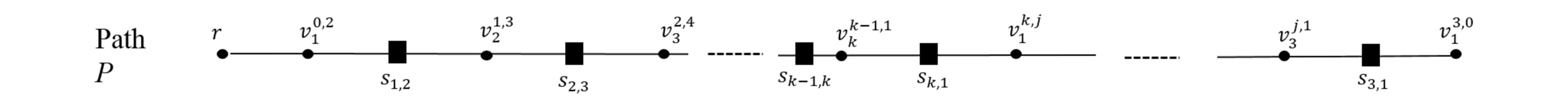}
  \caption{Path $P$ on $2 {k \choose 2}+2$ vertices. Selection vertices are depicted by squares.} \label{fig:main-c}
\end{figure}

The first $k$ selector vertices, namely $s_{1,2}, s_{2,3}, \ldots, s_{k-1,k}, s_{k,1}$ will play a special role that will become clear later.  To this end, for $1 \leq i \leq k$, we add an edge between the selector vertex $s_{i, i {\pmod k}+1}$ and each of the occurrences of $v_i$ that appear after $s_{i, i {\pmod k}+1}$ in $P$. These edges will be called the \emph{jumping edges} of $H$.

Let us denote by $F$ the graph constructed so far; see Fig.~\ref{fig:main-d}. Finally, in order to construct $H$, we replace each vertex of the form $v_j^{i,\ell}$ in $F$ with a whole copy of the vertex set $V_j$ of $G$ and make each of these new vertices adjacent to all the neighbors of $v_j^{i,\ell}$ in $F$. This completes the construction of $H$; see Fig.~\ref{fig:main-e}. Note that $\td(H) \leq {k \choose 2} + 1$, as the removal of the ${k \choose 2}$ selector vertices from $H$ results in a star centered at $r$ and isolated vertices.

\begin{figure}[t]
  \centering
  \hspace*{-2.8cm}\includegraphics[scale=0.13]{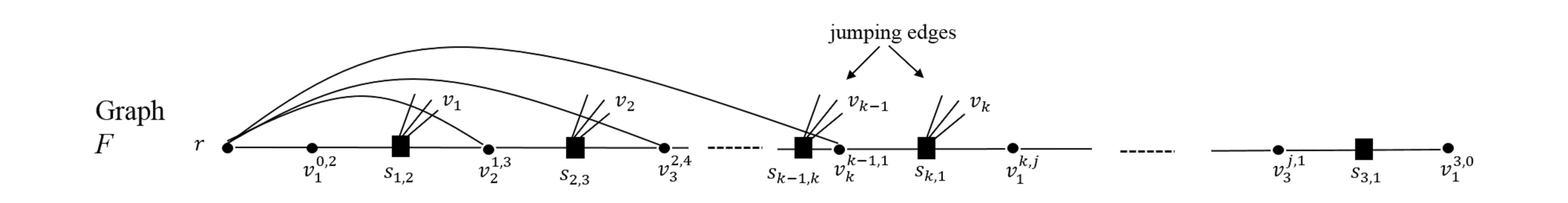}
  \caption{Graph $F$.} \label{fig:main-d}\vspace{-.1cm}
\end{figure}

\begin{figure}[h!]
  \centering
 \vspace{-.35cm}
 \hspace*{-3.0cm} \includegraphics[scale=0.13]{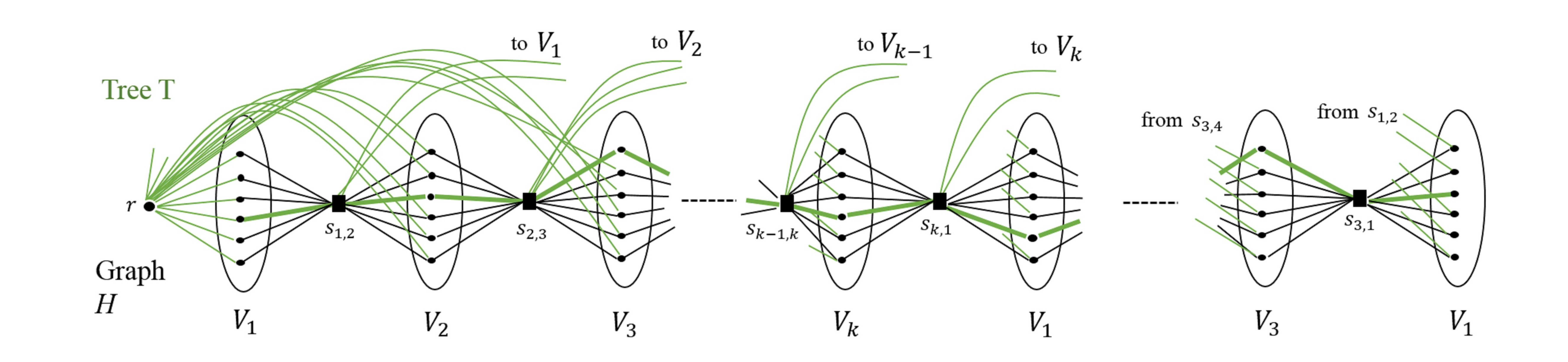}
  \caption{Graph $H$ and a solution arborescence $T$ drawn in green.} \label{fig:main-e}
\end{figure}

We now proceed to describe the color palette $X$, the coloring function $\chi$, and the cost function $\cc$, which altogether will encode the edges of $G$ and will ensure the desired properties of the reduction. For simplicity, we associate a distinct color with each edge of $H$, and thus, with slight abuse of notation, it is enough to describe the cost function $\cc$ for every ordered pair of incident edges of $H$. We will use just three different costs: $0$, $1$, and $B$, where $B$ can be set as any real number strictly greater than ${k \choose 2}$.  For each ordered pair of  incident edges $e_1,e_2$ of $H$, we define
\[\cc(e_1,e_2) =\left\{\begin{array}{lll}
& 0, & \mbox{if $e_1= \{\hat{x},s_{i,j}\}$ and $e_2 = \{s_{i,j},\hat{y}\}$ is a jumping edge such that}\\
 & & \mbox{$\hat{x},\hat{y}$ are copies of vertices $x,y \in V_i$, respectively, with $x \neq y$, or}\vspace{.15cm}\\
  & & \mbox{if $e_1=\{r,\hat{x}\}$ and $e_2=\{\hat{x},s_{1,2}\}$, where $\hat{x}$ is a copy of a vertex }\\
   & & \mbox{$x \in V_1$, or}\vspace{.15cm}\\

 & & \mbox{if $e_1$ and $e_2$ are the two edges that connect a vertex in a copy }\\
 & & \mbox{of a color class $V_i$ to a selector vertex.}\\  \vspace{-.2cm}
 & & \\
& 1, & \mbox{if $e_1= \{\hat{x},s_{i,j}\}$ and $e_2= \{s_{i,j},\hat{y}\}$, where $\hat{x}$ is a copy of a vertex}\\
& & \mbox{$x \in V_i$ and $\hat{y}$ is a copy of a vertex $y \in V_j$ such that $\{x,y\} \in E(G)$.}\\ \vspace{-.2cm}
 & & \\
& B, & \mbox{otherwise}.
\end{array}\right.\]

This completes the construction of $(H,X,\chi,r,\cc)$, which can be clearly performed in polynomial time.

\begin{claimN}\label{claim:bounded-average-degree}
The average degree of $H$ is bounded by $8$.
\end{claimN}
\begin{proofof}
Recall that we assumed that there are $k$ color classes, each with $n$ vertices. Note that there are $k \choose 2$ selector vertices and a root vertex $r$. Since each vertex of the form $v_j^{i,\ell}$ is replaced with a whole copy of the vertex set $V_j$, there are ${k \choose 2}+1 n$  additional vertices. Therefore, $|V(H)|=({k \choose 2}+1)n + {k \choose 2}+1$. Moreover, ${k \choose 2}-k$ selector vertices do not have any jumping edges; the neighborhood of such a selector vertex is exactly the vertices corresponding to its neighbors on the simple path $P$ described previously. Thus, the number of edges incident to these vertices without jumping edges is $({k \choose 2}-k)2n$. For each of the first $k$ selector vertices, when we sum up its jumping edges and its edges incident to the $n$ vertices corresponding to its successor vertex on path $P$, we get at most $(k-1)n$ edges because at most $k-1$ edges incident to the corresponding vertex remain on the Eulerian circuit. Together with its edges incident to the $n$ vertices corresponding to its predecessor vertex on path $P$, each of the first $k$ selector vertices has $k^2 n$ incident edges. Since root vertex $r$ is adjacent to the first occurrence of vertex $v_j$ in $P$ for $1 \le j \le k$, $kn$ edges are incident to vertex $r$. Hence, if we denote by $\overline{{\sf deg}}(H)$ the average degree of $H$, we get:
$$
\overline{{\sf deg}}(H)=\dfrac{2|E(H)|}{|V(H)|} \le \dfrac{kn+k^2 n + ({k \choose 2}-k)2n}{({k \choose 2}+1)n + {k \choose 2}+1} = \dfrac{8k (k-1)}{2k^2 -3k +4} \leq 8. \vspace{-.65cm}
$$
\end{proofof}

\vspace{.15cm}

We now claim that $H$ contains and arborescence $T$ rooted at $r$ with cost at most ${k \choose 2}$ if and only if $G$ contains a multicolored $k$-clique\footnote{If the costs associated with colors are restricted to be strictly positive, we can just replace cost 0 with cost $\varepsilon$, for an arbitrarily small positive real number $\varepsilon$, and ask for an arborescence in $H$ of cost strictly smaller than ${k \choose 2} + 1$.}. Note that the simple path $P$ described above naturally defines a partial left-to-right ordering among the vertices of $H$, and hence any arborescence rooted at $r$ contains \emph{forward} and \emph{backward} edges defined in an unambiguous way. Note also that all costs that involve a backward edge are equal to $B$, and therefore no such edge can be contained in an arborescence of cost at most ${k \choose 2}$.

Suppose first that $G$ contains a multicolored $k$-clique with vertices $v_1,v_2, \ldots, v_k$, where $v_i \in V_i$ for $1 \leq i \leq k$. Then we define the edges of the spanning tree $T$ of $H$ as follows.  Tree $T$ contains the edges of a left-to-right path $Q$ that starts at the root $r$, contains all ${k \choose 2}$ selector vertices and connects them, in each occurrence of a set $V_i$, to the copy of vertex $v_i$ defined by the $k$-clique. Since in $Q$ the selector vertices connect copies of pairwise adjacent vertices of $G$, the cost incurred so far by $T$ is exactly ${k \choose 2}$. For $1 \leq i \leq k$, we add to $Q$ the edges from $r$ to all vertices in the first occurrence of $V_i$ that are not contained in $Q$. Note that the addition of these edges to $T$ incurs no additional cost. Finally, we will use the jumping edges to reach the uncovered vertices of $H$. Namely, for $1 \leq i \leq k$, we add to $T$ an edge between the selector vertex $s_{i, i {\pmod k}+1}$ and all occurrences of the vertices in $V_i$ distinct from $v_i$ that appear after $s_{i, i {\pmod k}+1}$;  see the green  tree in Fig.~\ref{fig:main-e}. Note that since the jumping edges in $T$ contain copies of vertices distinct from the the ones in the $k$-clique, these edges incur no additional cost either. Therefore, $\cc(T,r) = {k \choose 2}$, as we wanted to prove.

Conversely, suppose now that $H$ has an arborescence $T$ rooted at $r$ with cost at most ${k \choose 2}$. Clearly, all costs incurred by the edges in $T$ are either 0 or 1. For a selector vertex $s_{i,j}$, we call the edges joining $s_{i,j}$ to the vertices in the occurrence of $V_i$ right before $s_{i,j}$ (resp., in the occurrence of $V_j$ right after $s_{i,j}$) the \emph{left} (resp., \emph{right}) edges of this selector vertex.

\begin{claimN}\label{claim:selection}
Tree $T$ contains exactly one left edge and exactly one right edge of  each selector vertex of $H$.
\end{claimN}
\begin{proofof}  Since only forward edges are allowed in $T$, and $T$ should be a tree, clearly for each selector vertex exactly one of its left edges belongs to $T$. Thus, it just remains to prove that  $T$ contains exactly one right edge of each selector vertex.

Let $s_{i,j}$ and $s_{j,\ell}$ be two consecutive selector vertices. Let $e$ be the left edge of $s_{j,\ell}$ in $T$ and let $v_j$ be the vertex of the copy of $V_j$ contained in $e$. Again, since backward edges are not allowed in $T$, $v_j$ needs to be incident with another forward edge $e'$ of $T$. If this edge $e'$ contains $r$ or if it is a jumping edge, then the cost incurred in $T$ by $e'$ and $e$ would be equal to $B$, a contradiction to the hypothesis that  $\cc(T,r) \leq {k \choose 2}$. Therefore, $e'$ is necessarily one of the right edges of $s_{i,j}$, so at least one of the right edges of selector vertex $s_{i,j}$ belongs to $T$.

As for the right edges of the last selector vertex, namely $s_{3,1}$, if none of them belonged to $T$, then there would be a jumping edge going to the last copy of $V_1$ such that, together with the left edge of selector vertex $s_{1,2}$ that belongs to $T$, would incur a cost of $B$, which is impossible.

We have already proved that exactly one left edge and {\sl at least} one of the right edges of  each selector vertex  belong to $T$. For each selector vertex $s_{i,j}$, its left edge in $T$ together with {\sl each} of its right edges in $T$ incur at cost of at least 1. But as there are ${k \choose 2}$ selector vertices in $H$, and by hypothesis the cost of $T$ is at most ${k \choose 2}$, we conclude that exactly one of the right edges of each selector vertex  belongs to $T$, as we wanted to prove.\end{proofof}

By Claim~\ref{claim:selection}, tree $T$ contains a path $Q'$ that chooses exactly one vertex from each occurrence of a color class of $G$. We shall now prove that, thanks to the jumping edges, these choices are \emph{coherent}, which will allow us to extract the desired multicolored $k$-clique in $G$.

\begin{claimN}\label{claim:coherence}
For every $1 \leq i \leq k$, the vertices in the copies of color class $V_i$ contained in $Q'$ all correspond to the same vertex of $G$, denoted by $v_i$.
\end{claimN}
\begin{proofof} Assume for contradiction that for some index $i$,  the vertices in the copies of color class $V_i$ contained in $Q'$ correspond to at least two distinct vertices $v_i$ and $v_i'$ of $G$, in such a way that $v_i$ is the selected vertex in the first occurrence of $V_i$, and $v_i'$ occurs later, say in the $j$th occurrence of $V_i$. Therefore, the copy of $v_i$ in the $j$th occurrence of $V_i$ does not belong to path $Q'$, so for this vertex to be contained in $T$, by construction it is necessarily an endpoint of a jumping edge $e$ starting at the selector vertex $s_{i, i {\pmod k}+1}$. But then the cost incurred in $T$ by the edges $e'$ and $e$, where $e'$ is the edge joining the copy of $v_i$ in the first occurrence of $V_i$ to the selector  vertex $s_{i, i {\pmod k}+1}$, equals $B$, contradicting the hypothesis that $\cc(T,r) \leq {k \choose 2}$.\end{proofof}

Finally, we claim that the vertices $v_1,v_2,\ldots,v_k$ defined by Claim~\ref{claim:coherence} induce a multicolored $k$-clique in $G$. Indeed, assume for contradiction that there exist two such vertices $v_i$ and $v_j$ such that $\{v_i,v_j\} \notin E(G)$. Then the cost in $T$ incurred by the two edges connecting the copies of $v_i$ and $v_j$ to the selector vertex $s_{i,j}$ (by Claim~\ref{claim:selection}, these two edges indeed belong to $T$) would be equal to $B$, contracting again the hypothesis that $\cc(T,r) \leq {k \choose 2}$. This concludes the proof of the theorem. \end{proof}

%\ig{Observation: the problem remains {\sc{W[1]}}-hard even restricted to graphs of bounded average degree, as the graph constructed above has average degree (at most) 8.}

%\begin{figure}[h!]
%\begin{center}
%\commentfig{
%%\vspace{-.2cm}
%\hspace*{-6cm}
%\includegraphics[width=1.35\textwidth angle=270]{reduction.pdf}%\vspace{-.5cm}

%\subfigure[caption]{\commentfig{\includegraphics[width=1.15\textwidth]{figures/figure-2a.pdf}}
%}\\
%\subfigure[caption]{\commentfig{\includegraphics[width=1.15\textwidth]{figures/figure-2b.pdf}}
%} \\
%\subfigure[caption]{\commentfig{\includegraphics[width=1.15\textwidth]{figures/figure-2c.pdf}}
%}\\
%\subfigure[caption]{\commentfig{\includegraphics[width=1.15\textwidth]{figures/figure-2d.pdf}}
%}\\
%\subfigure[caption]{\commentfig{\includegraphics[width=1.15\textwidth]{figures/figure-2e.pdf}}
%}\\

%}
%
%\caption{Example of the construction in the proof of Theorem~\ref{thm:hardness-treewidth}. \textbf{(a)} The complete graph $K_k$ and an Eulerian circuit $W$ in $K_k$ starting with $v_1,v_2,\ldots,v_k,v_1$ and ending with $v_3,v_1$. A $k$-colored graph $G$ is also illustrated. \textbf{(b)} A left-to-right representation of the Eulerian circuit $W$. \textbf{(c)} The path $P$ on $2 {k \choose 2}+2$ vertices. Selection vertices are depicted by squares. \textbf{(d)} The graph $F$. \textbf{(e)} The graph $H$ and a solution arborescence $T$ drawn in green. } \label{fig:reduction}
%\end{center}
%\end{figure}

%\subsection{W[1]-hardness on multigraphs with parameter tree-cut width}\label{sec:hard-tree-cut-width}
\newpage

In the next theorem we prove that the $\prbcc$ problem is {\sc{W[1]}}-hard on multigraphs parameterized by the tree-cutwidth of the input graph. Note that this result does not imply Theorem~\ref{thm:hardness-treewidth}, which applies to graphs without multiple edges.

\begin{theorem}\label{thm:hardness-treecutwidth}
The $\prbcc$ problem is {\sc{W[1]}}-hard on multigraphs parameterized by the tree-cutwidth of the input multigraph.
\end{theorem}
\begin{proof} As in Theorem~\ref{thm:hardness-treewidth}, we reduce again  from \textsc{Multicolored $k$-Clique}. Given an instance $(G,c,k)$ of \textsc{Multicolored $k$-Clique} with $k$ odd, we proceed to construct an instance $(H,X,\chi,r,\cc)$ of \prbcc. The first steps of the construction resemble the ones of Theorem~\ref{thm:hardness-treewidth}. Namely, let $F$ be the graph constructed in the proof of Theorem~\ref{thm:hardness-treewidth} (see Fig.~\ref{fig:main-d} for an illustration), and let $F'$ be the graph constructed from $F$ as follows. We delete the last vertex of $F$, namely $v_1^{3,0}$, and all edges incident with the root $r$ except the edge $\{r, v_1^{0,2}\}$. Finally, for every vertex of $F'$ of the form $v_{\ell}^{i,j}$ (that is, a vertex that is neither the root nor a selector vertex), let $e_1$ and $e_2$ be the two edges of the path $P$ incident with $v_{\ell}^{i,j}$, such that $e_1$ is to the left of $e_2$. Then we contract the edge $e_2$, and we give to the newly created vertex the name of the selector vertex incident with $e_2$. This completes the construction of $F'$. Note that $|V(F')| = {k \choose 2} + 1$. Finally, in order to construct $H$, we proceed as follows. For every edge $e$ of $F'$ which is not a jumping edge, let $s_{i,j}$ be its right endpoint. Then we replace $e$ with a multiedge with multiplicity $|V_i|$, and we associate each of these edges with a distinct vertex in $V_i \subseteq V(G)$. These edges are called the \emph{horizontal} edges of $H$. On the other hand, for every $1 \leq i \leq k$, and for every jumping edge $e$ whose left endpoint is the selector vertex $s_{i, i {\pmod k}+1}$, we replace $e$ with a multiedge with multiplicity $|V_i|$, and we subdivide each of these new edges once. Each of these new vertices $\hat{x}$ is associated with a distinct vertex in $V_i \subseteq V(G)$.  Let us call the selector vertices of the form $s_{i, i {\pmod k}+1}$ \emph{special} selector vertices. This completes the construction of $H$. Note that, as in the proof of Theorem~\ref{thm:hardness-treewidth}, the average degree of $H$ is also bounded by a constant (multiedges are counted with their multiplicity)

\begin{claimN}\label{claim:bounded-tcw}
The tree-cutwidth of $H$ is at most ${k \choose 2} + 1$.
\end{claimN}
\begin{proofof} We proceed to construct a tree-cut decomposition $(T, \mathcal{X})$ of $H$ of width at most ${k \choose 2} + 1$. Let $T$ be a star with $|V(H)| - {k \choose 2} - 1$ leaves rooted at its center. If $t$ is the center of this star $T$, then the bag $X_t$ contains the root $r$ of $H$ together with the ${k \choose 2}$ selector vertices. If $t$ is a leaf of $T$, then the bag $X_t$ contains a single vertex, in such a way that each of the remaining $|V(H)| - {k \choose 2} - 1$ vertices of $T$ is associated with one of the leaves. For every leaf $t \in V(T)$, it holds that $\adh(t) = 2$, as every vertex in $H$ that is neither the root nor a selector vertex has degree exactly 2.
%Let $H_t$ be the torso of $H$ at a vertex $t \in V(T)$, and let $\bar{H}_t$ be its corresponding $3$-center.
Also, for every leaf $t$ of $T$, clearly $\tor(t) \leq 2$, as $|X_t| = 1$ and $t$ has degree 1 in $T$. Finally, if $t$ the root of $T$, then when considering the torso $H_t$, every vertex in a leaf-bag gets dissolved, as each such vertex has exactly 2 neighbors in $X_t$. Therefore, $\tor(t) \leq  {k \choose 2} + 1$.
\end{proofof}

We now proceed to describe the color palette $X$, the coloring function $\chi$, and the cost function $\cc$, which altogether will encode the edges of $G$ and will ensure the desired properties of the reduction. For simplicity, as in the proof of Theorem~\ref{thm:hardness-treewidth}, we again associate a distinct color with every edge of $H$, and thus, it is enough to describe the cost function $\cc$ for every ordered pair of incident edges of $H$. In this case, we will use just two different costs: $0$ and  $1$.  For every ordered pair of  incident edges $e_1,e_2$ of $H$, we define
\[\cc(e_1,e_2) =\left\{\begin{array}{lll}
& 0, & \mbox{if $e_1= \{x, s_{i,j}\}$ and $e_2= \{s_{i,j}, y\}$ are two horizontal edges}\\
& & \mbox{such that $x$ is to the left of $y$, and the vertex in $V_i$ associated}\\
& & \mbox{with $e_1$ is adjacent in $G$ to the vertex in $V_j$ associated with $e_2$, or} \vspace{.15cm}\\

  & & \mbox{if $e_1 = \{x, s_{i, i{\pmod k}+1}\}$ and $e_2 = \{s_{i, i{\pmod k}+1}, \hat{x}_i'\}$ are}\\
   & & \mbox{such that $ s_{i, i {\pmod k}+1}$ is a special selector vertex,  edge $e_1$}\\
   & & \mbox{is horizontal and is associated with a vertex $v_i \in V_i$, edge $e_2$}\\
   & & \mbox{arises from the subdivision of a jumping edge such that vertex}\\
   & & \mbox{$\hat{x}_i'$ is associated with a vertex $v_i' \in V_i$ with $v_i \neq v_i'$, or }
   \vspace{.15cm}\\

  & & \mbox{if $e_1 = \{x, s_{i, j}\}$ and $e_2 = \{s_{i, j}, \hat{x}_i'\}$ with $ s_{i, j}$ being a selector}\\
   & & \mbox{vertex that is not special,  edge $e_1$ is horizontal and is associated}\\
   & & \mbox{with a vertex $v_i \in V_i$, edge $e_2$ arises from the subdivision of a}\\
   & & \mbox{jumping edge such that vertex $\hat{x}_i'$ is associated with a vertex}\\
   & & \mbox{$v_i' \in V_i$ with $v_i = v_i'$.}\\

 \vspace{-.2cm}
 & & \\
& 1, & \mbox{otherwise}.
\end{array}\right.\]

The three different cases above where $\cc(e_1,e_2)=0$ are illustrated in Fig~\ref{fig:reduction-tcw}(a)-(b)-(c), respectively. This completes the construction of $(H,X,\chi,r,\cc)$, which can be clearly performed in polynomial time. We now claim that $H$ contains and arborescence $T$ rooted at $r$ with cost $0$ if and only if $G$ contains a multicolored $k$-clique. Again, we assume that any arborescence in $H$ rooted at $r$ contains \emph{forward} and \emph{backward} edges defined in an unambiguous way.

\begin{figure}[htb]
\begin{center}
\commentfig{
\vspace{-.75cm}
\hspace*{-1.75cm}
\includegraphics[width=1.26\textwidth]{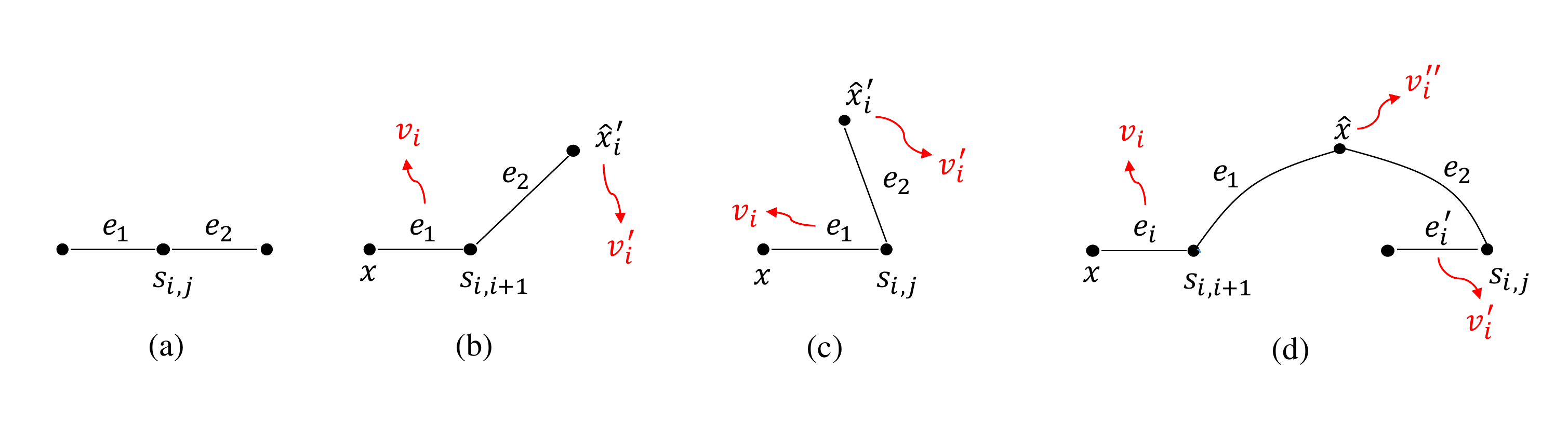}%\vspace{-.5cm}
}%\vspace{-2.5cm}
\caption{(a)-(b)-(c) The three cases where $\cc(e_1,e_2)=0$ in the proof of Theorem~\ref{thm:hardness-treecutwidth}. (d) Construction in the proof. The red vertices in the figure correspond to the vertices in $V_i$ that are associated with the corresponding edges or vertices of $H$.} \label{fig:reduction-tcw}
\end{center}
\end{figure}

%\begin{figure}[h!]
%\begin{center}
%\commentfig{
%%\vspace{-.2cm}
%%\hspace*{-6cm}
%\includegraphics[width=1.20\textwidth]{figures/figure-6.pdf}%\vspace{-.5cm}
%}
%\caption{(a)-(b)-(c) The three cases where $\cc(e_1,e_2)=1$ in the proof of Theorem~\ref{thm:hardness-treecutwidth}. (d) Construction in the proof. The red vertices in the figure correspond to the vertices in $V_i$ that are associated with the corresponding edge or vertex of $H$.} \label{fig:reduction-tcw}
%\end{center}
%\end{figure}

Suppose first that $G$ contains a multicolored $k$-clique with vertices $v_1,v_2, \ldots, v_k$, where $v_i \in V_i$ for $1 \leq i \leq k$. Then we define the edges of the spanning tree $T$ of $H$ as follows.  For each selector vertex $s_{i,j}$, $1 \leq i,j \leq k$, we add to $T$ its left horizontal edge associated with the vertex $v_i$ that belongs to the clique. For every jumping edge $\{s_{i, i{\pmod k}+1}, s_{i,j}\}$ of $T'$,
%for some $1 \leq j \leq k$ with $j \notin \{i , i{\pmod k}+1\}$,
we do the following. Note that this edge has given rise to $|V_i|$ paths with two edges in $H$, and the vertices of $H$ in the middle of these paths, which we call \emph{inner} vertices, are associated with the vertices in $V_i$. Then we add to $H$  a forward edge between $s_{i, i{\pmod k}+1}$ and each inner vertex $\hat{x}$ associated with a vertex in $V_i$ distinct from the vertex $v_i$ that belongs to the clique. Note that $|V_i| - 1$ edges are added to $H$ in this way. Finally, we add a backward edge between $s_{i,j}$ and the inner vertex $\hat{x}$ associated with the vertex $v_i \in V_i$ that belongs to the clique. By the definition of the cost function and using the fact that the vertices $v_1,v_2, \ldots, v_k$ are pairwise adjacent in $G$, it can be easily checked that $\cc(T,r) = 0$, as we wanted to prove.

Conversely, suppose now that $H$ has an arborescence $T$ rooted at $r$ with cost  $0$. Clearly, all costs incurred by the edges in $T$ are necessarily $0$. Since the cost incurred by the two edges incident with every inner vertex is equal to 1, necessarily $T$ contains a path $Q$ starting at the root $r$ and containing all ${k \choose 2}$ selector vertices. Let $s_{i,j}$ be an arbitrary selector vertex distinct from the last one, and let $e_i$ and $e_j$ be its left and right incident horizontal edges in $Q$, respectively. Since $\cc(e_1,e_2)=0$, necessarily the vertex $v_i \in V_i$ associated with $e_i$ is adjacent in $G$ to the vertex $v_j \in V_j$ associated with $e_j$. We say that the selector vertex $s_{i,j}$ has \emph{selected} the vertex $v_i$. Our objective is to prove that these selections are \emph{coherent}, in the sense that if two distinct selector vertices $s_{i,j}$ and $s_{i,\ell}$ have selected vertices $v_i$ and $v_i'$ in $V_i$, respectively, then $v_i = v_i '$. This property will be guaranteed by how the inner vertices are covered by $H$, as we proceed to prove.

By construction of $H$, each such inner vertex $\hat{x}$ is adjacent to a special selector vertex, say $s_{i,i+1}$, and to another selector vertex that is not special, say $s_{i,j}$. Let $e_1 = \{s_{i,i+1}, \hat{x} \}$ and let $e_2 = \{\hat{x}, s_{i,j} \}$. Note that either $e_1$ or $e_2$ belong to $T$, but not both, as otherwise these two edges would close a cycle with the path $Q$. Let also $e_i$ (resp., $e_i'$) be the edge that is to the left of $s_{i,i+1}$ (resp., $s_{i,j})$ in $Q$; see Fig~\ref{fig:reduction-tcw}(d) for an illustration. Note that $e_i$ (resp., $e_i'$) is associated with a vertex $v_i \in V_i$ (resp., $v_i' \in V_i$), and similarly vertex $\hat{x}$ is associated with another vertex $v_i'' \in V_i$. We distinguish two cases. Assume first that $v_i = v_i''$. In this case, by the definition of the cost function we have that $cc(e_i,e_1) = 1$, so $e_1$ cannot belong to $T$, implying that $e_2$ belongs to $T$, which is possible only if $cc(e_i',e_2) = 1$, and this is true if and only if $v_i' = v_i$. This implies that the selections made by the selection vertices are coherent, as we wanted to prove.  Otherwise, we have that $v_i \neq v_i''$. By the definition of the cost function, it holds that $cc(e_i',e_2) = 1$, and therefore, as we assume that $\cc(T,r) = 0$, necessarily $e_1$ belongs to $T$, which is indeed possible as  $cc(e_i,e_1) = 0$ because $v_i \neq v_i''$.

By the above discussion, it follows that for each $1 \leq i \leq k$, all selector vertices of the form $s_{i,j}$, for every $1 \leq j \leq k$, $i \neq j$, have selected the same vertex $v_i \in V_i$. Furthermore, for every $1 \leq i,j \leq k$, $i \neq j$, it holds that $\{v_i, v_j\} \in E(G)$. That is, the selected vertices $v_1,v_2,\ldots,v_k$  induce a multicolored $k$-clique in $G$, concluding the proof of the theorem.\end{proof}

\subsection{NP-hardness on planar graphs}\label{sec:hard-planar}
\begin{figure}[h!]
\centering
\vspace{1cm}
\vspace{-1.7cm}
  \includegraphics[scale=0.55]{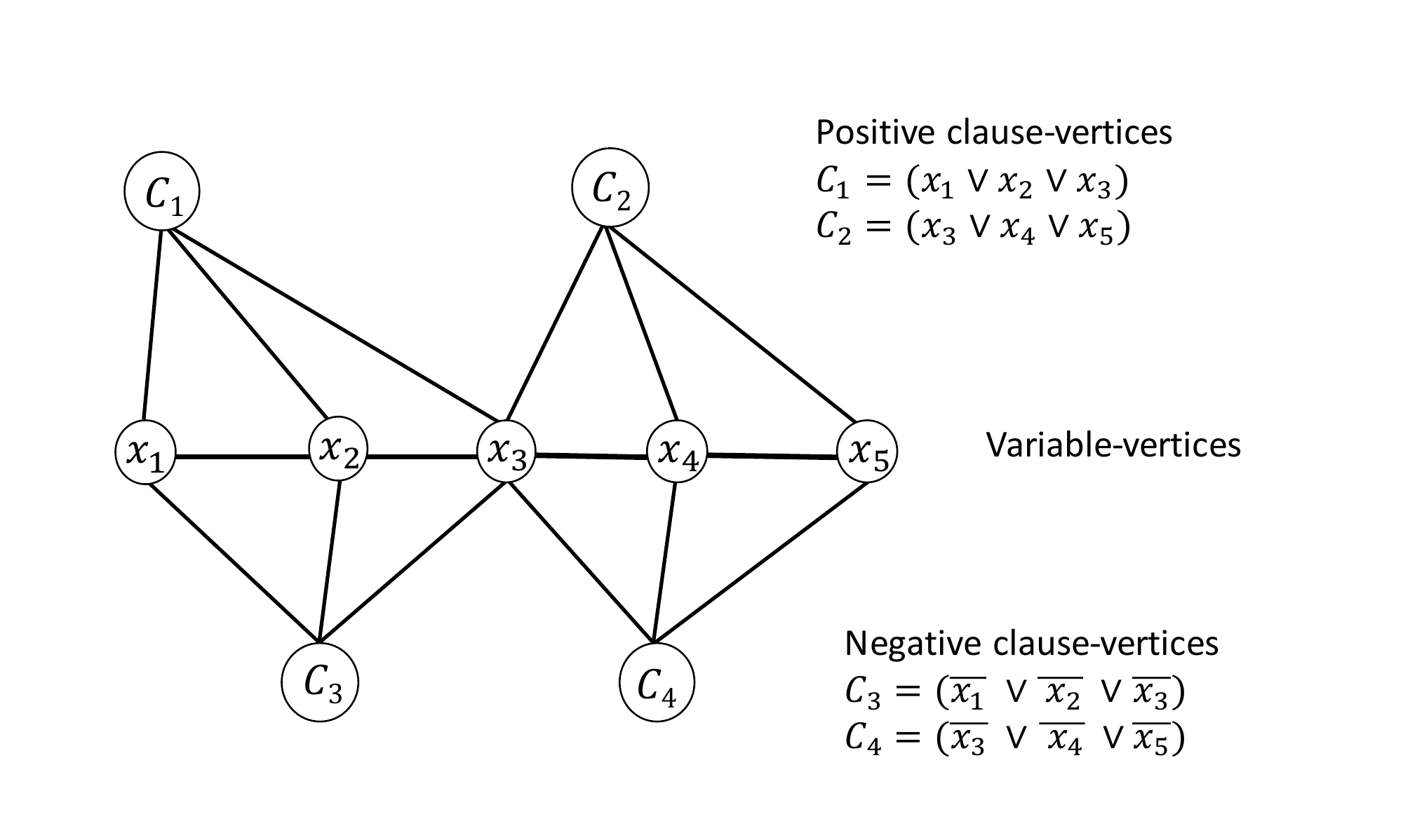}
 % \vspace{-2cm}
  \caption{A monotone rectilinear representation of a planar monotone \textsc{3-sat} instance.}
  \label{fig:planar3satexample}
\end{figure}

In this subsection we prove that the $\prbcc$ problem remains {\sf NP}-hard on planar graphs.  In order to prove this result, we need to introduce the \textsc{Planar Monotone 3-sat} problem. An
instance of \textsc{3-sat} is called \emph{monotone} if each clause is monotone, that is, each clause consists only of positive variables or only of negative variables.  We call a clause with only positive (resp., negative) variables
a \emph{positive} (resp., \emph{negative}) clause. Given an instance $\phi$ of \textsc{3-sat}, we define the bipartite graph $G_{\phi}$ that has one vertex per each variable and each clause, and has an edge between a variable-vertex and a clause-vertex if and only if the variable appears (positively or negatively) in the clause. A \emph{monotone rectilinear representation} of a monotone \textsc{3-sat} instance $\phi$ is a {\sl planar} drawing of $G_{\phi}$ such that all variable-vertices lie on a path, all positive clause-vertices lie {\sl above} the path, and all negative clause-vertices lie {\sl below} the path; see Fig.~\ref{fig:planar3satexample}.

In the \textsc{Planar Monotone 3-sat} problem, we are given a monotone rectilinear representation of a planar monotone \textsc{3-sat}
instance $\phi$, and the objective is to determine whether $\phi$ is satisfiable. Berg and Khosravi~\cite{BergK12} proved that the \textsc{Planar Monotone 3-sat} problem is {\sf NP}-complete.

\begin{theorem}\label{thm:hardness-planar}
The $\prbcc$ problem is {\sf NP}-hard on planar graphs even when restricted to instances with at most 6 colors and 0/1 symmetric costs.
\end{theorem}
\begin{proof} We reduce from the \textsc{Planar Monotone 3-sat} problem. Given a monotone rectilinear representation of a planar monotone \text{3-sat}
instance $\phi$, we build an instance $(H,X,\chi,r,f)$ of $\prbcc$ as follows. We denote the variable-vertices of $G_{\phi}$ as $\{x_1,\ldots,x_n\}$ and the clause-vertices of $G_{\phi}$ as $\{C_1,\ldots,C_m\}$. Without loss of generality, we assume that the variable-vertices appear in the order $x_1,\ldots,x_n$ on the path $P$ of $G_{\phi}$ that links the variable-vertices. For every variable-vertex $x_i$ of $G_{\phi}$, we add to $H$ a gadget consisting of four vertices $x_i^{\ell},x_i^{r},x_i^{+},x_i^{-}$ and five edges $\{x_i^{\ell}, x_i^{+}\}$, $\{x_i^{+},x_i^{r}\}$, $\{x_i^{r},x_i^{-}\}$, $\{x_i^{-},x_i^{\ell}\}$, $\{x_i^{+},x_i^{-}\}$. We add to $H$ a new vertex $r$, which we set as the root, and we add the edge $\{r,x_1^{\ell}\}$. For every $i \in \{1,\ldots,n-1\}$, we add to $H$ the edge $\{x_i^r,x_{i+1}^{\ell}\}$. We add to $H$ all clause-vertices $C_1,\ldots,C_m$. For every $i \in \{1,\ldots,n\}$, we add an edge between vertex $x_i^+$ and each clause-vertex of $G_{\phi}$ in which variable $x_i$ appears positively, and an  edge between vertex $x_i^-$ and each clause-vertex of $G_{\phi}$ in which variable $x_i$ appears negatively. This completes the construction of $H$, which is illustrated in Fig.~\ref{fig:reduction-planar}. Since $G_{\phi}$ is planar and all positive (resp., negative) clause-vertices appear above (resp., below) the path $P$, it is easy to see that the graph $H$ is planar as well.

\begin{figure}[t]
\begin{center}
\commentfig{
\vspace{-0.8cm}
\hspace*{-2.5cm}
\includegraphics[width=1.5\textwidth,  angle =0]{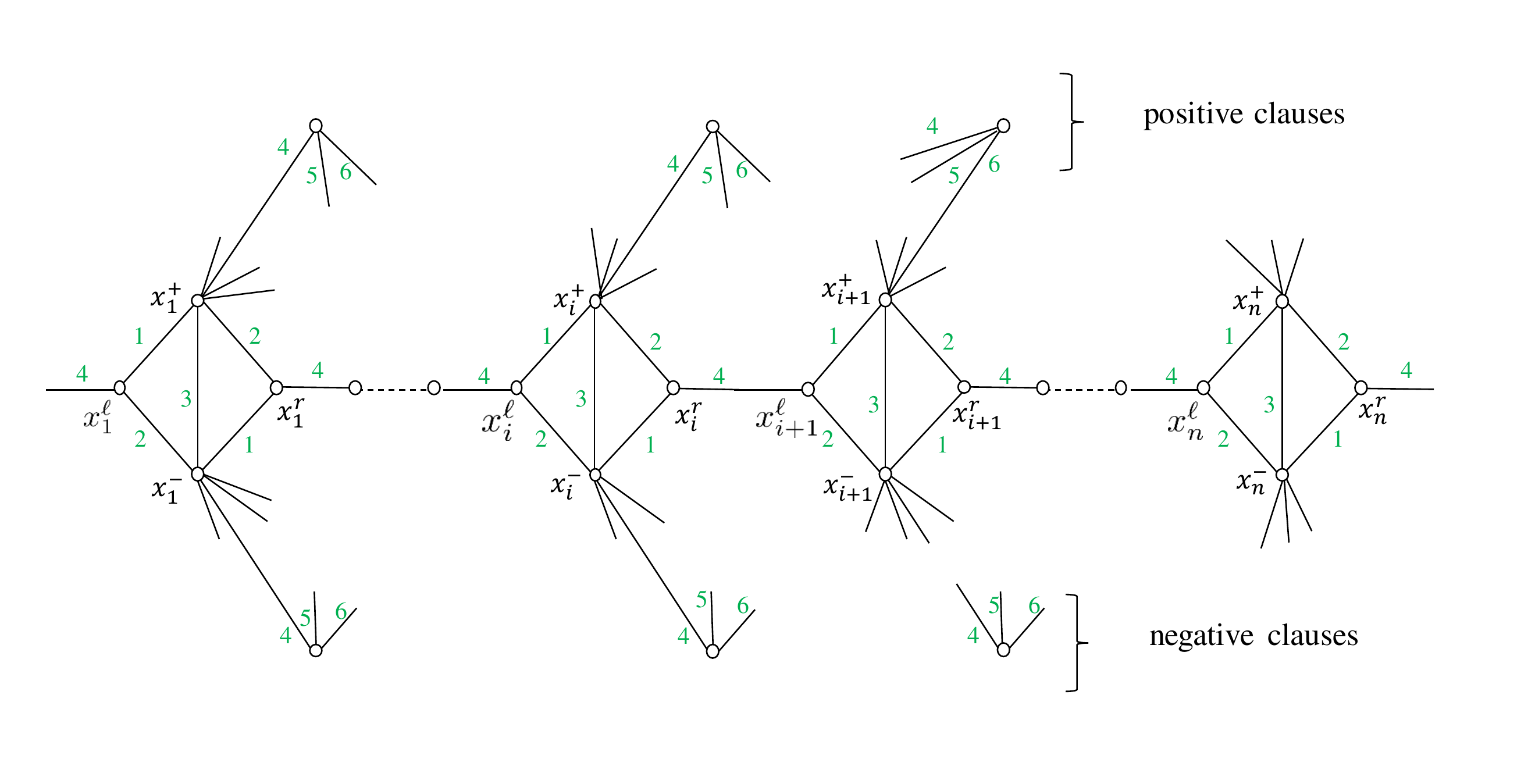}\vspace{2cm}
}\vspace{-2.7cm}
\caption{The graph $H$ constructed in the proof of Theorem~\ref{thm:hardness-planar}, together with  the edge-coloring function $\chi$ (in green). } \label{fig:reduction-planar}
\end{center}
\end{figure}

We define the color palette as $X=\{1,2,3,4,5,6\}$. Let us now describe the edge-coloring function $\chi$. For every clause-vertex $C_j$, we color arbitrarily its three incident edges with the colors $\{4,5,6\}$, so that each edge incident to $C_j$ gets a different color. For every $i \in \{1,\ldots,n\}$, we define $\chi(\{x_i^{\ell},x_i^+\}) = \chi(\{x_i^{r},x_i^-\}) = 1$, $\chi(\{x_i^{+},x_i^r\}) = \chi(\{x_i^{-},x_i^{\ell}\}) = 2$, and $\chi(\{x_i^{+},x_i^-\})=3$. We set $\chi(\{r,x_1^{\ell}\})=4$ and for every $i \in \{1,\ldots,n-1\}$, $\chi(\{x_i^{r},x_{i+1}^{\ell}\})=4$. The function $\chi$ is also depicted in Fig.~\ref{fig:reduction-planar}. Finally, we define the cost function $\cc$ to be symmetric and, for every $i \in \{1,2,3,4,5,6\}$, we set $\cc(i,i)=0$. We define $\cc(1,2)=1$ and $\cc(1,3)=\cc(2,3)=0$. For every $i \in \{4,5,6\}$, we set $\cc(1,i)= \cc(2,i) = 0$ and $\cc(3,i) = 1$. Finally, for every $i,j \in \{4,5,6\}$ with $i \neq j$ we set $\cc(i,j)=1$.

We now claim that $H$ contains an arborescence $T$ rooted at $r$ with cost 0 if and only if the formula $\phi$ is satisfiable.

Suppose first that $\phi$ is satisfiable, and fix a satisfying assignment of $\phi$. We proceed to define an arborescence $T$ rooted at $T$ with cost 0. $T$ contains the edge $\{r, x_1^{\ell}\}$ and, for every $i \in \{1,\ldots,n-1\}$, the edge $\{x_i^r,x_{i+1}^{\ell}\}$. For every $i \in \{1,\ldots,n\}$, if variable $x_i$ is set to 1 in the satisfying assignment of $\phi$, we add to $T$ the edges $\{x_i^{\ell}$, $x_i^{+}\}$, $\{x_i^{+},x_i^{-}\}$, and $\{x_i^{-},x_i^{r}\}$. Otherwise, if variable $x_i$ is set to 0 in the satisfying assignment of $\phi$, we add to $T$ the edges $\{x_i^{\ell}$, $x_i^{-}\}$, $\{x_i^{-},x_i^{+}\}$, and $\{x_i^{+},x_i^{r}\}$. Finally, for every $j \in \{1,\ldots,m\}$, let $x_t$ (resp., $\bar{x}_t$) be a literal in clause $C_j$ that is set to 1 by the satisfying assignment of $\phi$ (note that for each clause we consider only {\sl one} such literal). Then we add to $T$ an edge between vertex $C_j$ and vertex $x_t^+$ (resp., $x_t^-$). It can be easily checked that $T$ is an arborescence of $H$ with cost 0.

Conversely, suppose now that $H$ contains an arborescence $T$ rooted at $r$ with cost 0, and let us define a satisfying assignment of $\phi$. Since for every $i,j \in \{4,5,6\}$ with $i \neq j$ we have that $\cc(i,j)=1$, for every clause-vertex $C_j$ exactly one of its incident edges belongs to $T$. From the structure of $H$ and from the fact that $\cc(1,2)=1$, it follows that in order for the tree $T$ to span all vertices of $H$, for every $i \in \{1,\ldots,n\}$ either the three edges $\{x_i^{\ell}$, $x_i^{+}\}$, $\{x_i^{+},x_i^{-}\}$, $\{x_i^{-},x_i^{r}\}$ or the three edges $\{x_i^{\ell}$, $x_i^{-}\}$, $\{x_i^{-},x_i^{+}\}$, $\{x_i^{+},x_i^{r}\}$ belong to $T$. In the former case, we set variable $x_i$ to 1, and in the latter case we set variable $x_i$ to 0. Since for every $i \in \{4,5,6\}$ we have that $\cc(3,i)=1$, it follows that for every clause-vertex $C_j$, its incident edge that belongs to $T$ joins $C_j$ to a literal that is set to 1 by the constructed assignment. Hence, all clauses of $\phi$ are satisfied by this assignment, concluding the proof of the theorem. \end{proof}

%\ig{we can bound the maximum degree of the graph built in the above reduction by allowing non-symmetric costs}
%\textcolor{blue}{TODO by DIDEM}

%\begin{figure}[h!]
%\begin{center}
%\commentfig{
%\vspace{-3cm}
%\hspace*{-2.5cm}
%\includegraphics[width=1.5\textwidth,  angle =0]{figures/Figure-5.pdf}\vspace{2cm}
%}
%\caption{The graph $H$ constructed in the proof of Theorem~\ref{thm:hardness-planar-boundeddegree}, together with  the edge-coloring function $\chi$.} \label{fig:reduction-planar-boundeddegree}
%\end{center}
%\end{figure}

Note that the above proof actually implies that $\prbcc$ cannot be approximated to {\sl any} positive ratio on planar graphs in polynomial time, since an optimal solution has cost $0$. We do not know whether such a strong inapproximability result holds even if we do not allow to use costs $0$ among different colors.

%Can we transform this to an inapproximability result without using such costs

In the next theorem we present a modification of the previous reduction showing that the $\prbcc$ problem remains hard even if the maximum degree of the input planar graph is bounded.

\begin{theorem}\label{thm:hardness-planar-boundeddegree}
The $\prbcc$ problem is {\sf NP}-hard on planar graphs even when restricted to instances with at most 8 colors, maximum degree bounded by $4$, and 0/1 symmetric costs.
\end{theorem}
\begin{proof} The reduction follows closely the one of Theorem~\ref{thm:hardness-planar}. Given a monotone rectilinear representation of a planar monotone \textsc{3-sat} instance $\phi$, we build an instance $(H,X,\chi,r,f)$ of $\prbcc$ as follows. We denote the variable-vertices of $G_{\phi}$ as $\{x_1,\ldots,x_n\}$ and the clause-vertices of $G_{\phi}$ as $\{C_1,\ldots,C_m\}$. Without loss of generality, we assume that the variable-vertices appear in the order $x_1,\ldots,x_n$. For every variable-vertex $x_i$ of $G_{\phi}$, we add to $H$ a gadget consisting of four vertices $x_i^{\ell},x_i^{r},x_i^{+},x_i^{-}$ and five edges $\{x_i^{\ell}, x_i^{+}\}$, $\{x_i^{+},x_i^{r}\}$, $\{x_i^{r},x_i^{-}\}$, $\{x_i^{-},x_i^{\ell}\}$, $\{x_i^{+},x_i^{-}\}$. We add to $H$ a new vertex $r$, which we set as the root, and we add the edge $\{r,x_1^{\ell}\}$. Let $\mathcal{C}_{i}^{+}$ be the set of clauses that variable $i$ appears positively and let $\mathcal{C}_{i}^{-}$ be the set of clauses that variable $i$ appears negatively. For every $i \in \{1,\ldots,n\}$ and for every clause $j \in \set{1, \ldots, |\mathcal{C}_{i}^{+}|}$, we add vertices $x_{ij}^{+}$ and $x_{ij}^{r+}$. Likewise, for every $i \in \{1,\ldots,n\}$ and for every clause $j \in \set{1, \ldots, |\mathcal{C}_{i}^{-}|}$, we add vertices $x_{ij}^{-}$ and $x_{ij}^{r-}$. Moreover, for every $i \in \{1,\ldots,n-1\}$, we add a vertex $x_{i}'$ as well as the edges $\set{x_{i}^{r}, x_{i}'}$ and $\set{x_{i}', x_{i+1}^{l}}$. We proceed our construction by adding for every $i \in \{1,\ldots,n\}$ and $j \in \set{1, \ldots, |\mathcal{C}_{i}^{+}|}$ the edges $\set{x_{i}^{+}, x_{ij}^{+}}$, $\set{x_{i}^{r}, x_{ij}^{r+}}$, $\set{x_{ij}^{+}, x_{ij}^{r+}}$, $\set{x_{ij}^{+}, C_j}$
and for every $j \in \set{1, \ldots, |\mathcal{C}_{i}^{-}|}$ the edges $\set{x_{i}^{-}, x_{ij}^{-}}$, $\set{x_{i}^{'}, x_{ij}^{r-}}$, $\set{x_{ij}^{-}, x_{ij}^{r-}}$, $\set{x_{ij}^{-}, C_j}$. Subsequently, for every $i \in \{1,\ldots,n\}$ and $j \in \set{1, \ldots, |\mathcal{C}_{i}^{+}|-1}$ we add the edges $\set{x_{ij}^{+}, x_{i(j+1)}^{+}}$, $\set{x_{ij}^{r+}, x_{i(j+1)}^{r+}}$, and for every $j \in \set{1, \ldots, |\mathcal{C}_{i}^{-}|-1}$ we add the edges $\set{x_{ij}^{-}, x_{i(j+1)}^{-}}$, $\set{x_{ij}^{r-}, x_{i(j+1)}^{r-}}$. Note that the maximum degree of $H$ is indeed  4. An example of this construction can be found in Fig~\ref{fig:reduction-planar-boundeddegree}.

\begin{figure}[th]
\begin{center}
\commentfig{
\vspace{-.7cm}
\hspace*{-2.5cm}
\includegraphics[width=1.5\textwidth,  angle =0]{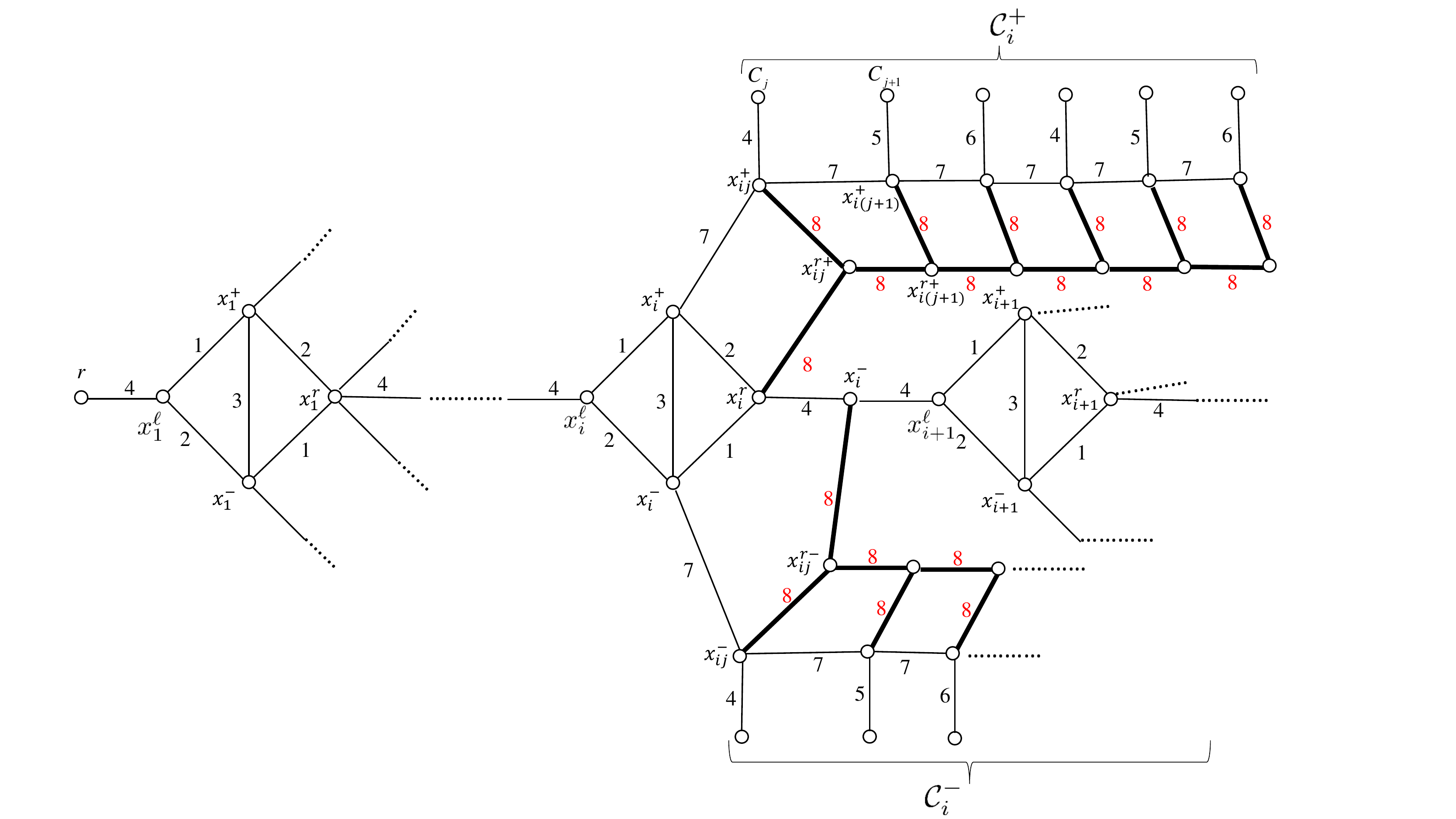}\vspace{2cm}
}\vspace{-2.1cm}
\caption{The graph $H$ constructed in the proof of Theorem~\ref{thm:hardness-planar-boundeddegree}, together with  the edge-coloring function $\chi$. Note that the maximum degree of $H$ is 4.} \label{fig:reduction-planar-boundeddegree}
\end{center}
\end{figure}

We define the color palette as $X=\set{1,2,3,4,5,6,7,8}$. Let us now describe the edge coloring function $\chi$. For every clause vertex, we arbitrarily color its three incident edges with colors $\set{4,5,6}$ so that each incident edge gets a different color. For every $i \in \{1,\ldots,n\}$, we define $\chi(\{x_i^{\ell},x_i^+\}) = \chi(\{x_i^{r},x_i^-\}) = 1$, $\chi(\{x_i^{+},x_i^r\}) = \chi(\{x_i^{-},x_i^{\ell}\}) = 2$, and $\chi(\{x_i^{+},x_i^-\})=3$. We set $\chi(\{r,x_1^{\ell}\})=4$ and for every $i \in \{1,\ldots,n-1\}$, $\chi(\{x_i^{r},x_{i}^{'})=\chi(x_{i}^{'},x_{i+1}^{l})=4$. For every $j\in \set{1, \ldots, |\mathcal{C}_{i}^{+}|-1}$, we set $\chi(x_{ij}^{+}, x_{i(j+1)}^{+})=7$ and $\chi(x_{ij}^{r+}, x_{i(j+1)}^{r+})=8$. Likewise, for every $j\in \set{1, \ldots, |\mathcal{C}_{i}^{-}|-1}$, we set $\chi(x_{ij}^{-}, x_{i(j+1)}^{-})=7$ and $\chi(x_{ij}^{r-}, x_{i(j+1)}^{r-})=8$. For every $i \in \{1,\ldots,n\}$ and $j\in \set{1, \ldots, |\mathcal{C}_{i}^{+}|}$, we set $\chi(x_{ij}^{+}, x_{ij}^{r+})=\chi(x_{i}^{r}, x_{ij}^{r+})=8$, $\chi(x_{i}^{+}, x_{ij}^{+})=7$ and for every $j\in \set{1, \ldots, |\mathcal{C}_{i}^{-}|}$ we set $\chi(x_{i}^{'}, x_{ij}^{r-})=\chi(x_{ij}^{-}, x_{ij}^{r-})=8$, $\chi(x_{i}^{-}, x_{ij}^{-})=7$. The function $\chi$ is also depicted in Fig.~\ref{fig:reduction-planar-boundeddegree}. Finally, we define the cost function $\cc$ as follows: for every $i \in \{1,2,3,4,5,6,7,8\}$, we set $\cc(i,i)=0$, $\cc(1,2)=1$ and $\cc(1,3)=\cc(2,3)=0$. For every $i,j \in \{4,5,6\}$ with $i \neq j$ we set $\cc(i,j)=1$. For all $i \in \set{1,2,3,4,5,6}$, $\cc(8,i)=0$, whereas $\cc(1,8)=\cc(2,8)=0$ and $\cc(7,8)=\cc(8,7)=1$. Moreover, $\cc(1,4)=\cc(2,4)=0$ and for all $i \in \set{4,5,6}$ we set $\cc(8,i)=\cc(i,8)=0$ and $\cc(7,i)=\cc(i,7)=1$.

We now claim that $H$ contains an arborescence $T$ rooted at $r$ with cost  0 if and only if the formula $\phi$ is satisfiable.

Suppose first that $\phi$ is satisfiable, and we proceed to define an arborescence $T$ rooted at $r$ with cost 0. $T$ contains the edge $\{r, x_1^{\ell}\}$ and, for every $i \in \set{1,\ldots,n-1}$, the edges $\set{x_{i}^{r}, x_{i}^{'}}$ and $\set{x_{i}^{'}, x_{i+1}^{\ell}}$. For every $i \in \set{1,\ldots,n}$, if variable $x_i$ is set to 1 in the satisfying assignment of $\phi$, we add to $T$ the edges $\{x_i^{\ell}$, $x_i^{+}\}$, $\{x_i^{+},x_i^{-}\}$, and $\{x_i^{-},x_i^{r}\}$. Otherwise, if variable $x_i$ is set to 0 in the satisfying assignment of $\phi$, we add to $T$ the edges $\{x_i^{\ell}$, $x_i^{-}\}$, $\{x_i^{-},x_i^{+}\}$, and $\{x_i^{+},x_i^{r}\}$. For every $i \in \set{1,\ldots,n}$, if variable $x_i$ is set to 1 in the satisfying assignment of $\phi$, then for every $j \in \set{1,\ldots, |C_{i}^{+}|}$ we add to $T$ the edges $\set{x_{i}^{r}, x_{ij}^{r+}}$, $\set{x_{ij}^{r+}, x_{ij}^{+}}$, $\set{x_{ij}^{+}, C_j}$ (note that for each clause $C_j$ we add only one such edge), for every $j \in \set{1,\ldots, |C_{i}^{+}|-1}$ we add to $T$ the edges $\set{x_{ij}^{r+}, x_{i(j+1)}^{r+}}$, for every $j \in \set{1,\ldots, |C_{i}^{-}|}$ we add to $T$ the edges $\set{x_{i}^{-}, x_{ij}^{-}}$, $\set{x_{i}^{'}, x_{ij}^{r-}}$, and for every $j \in \set{1,\ldots, |C_{i}^{-}|-1}$ we add to $T$ the edges $\set{x_{ij}^{r-}, x_{i(j+1)}^{r-}}$, $\set{x_{ij}^{-}, x_{i(j+1)}^{-}}$. For every $i \in \set{1,\ldots,n}$, if variable $x_i$ is set to 0 in the satisfying assignment of $\phi$, then for every $j \in \set{1,\ldots, |C_{i}^{+}|}$ we add to $T$ the edges $\set{x_{i}^{r}, x_{ij}^{r+}}$, $\set{x_{i}^{+}, x_{ij}^{+}}$, for every $j \in \set{1,\ldots, |C_{i}^{+}|-1}$ we add to $T$ the edges $\set{x_{ij}^{r+}, x_{i(j+1)}^{r+}}$, $\set{x_{ij}^{+}, x_{i(j+1)}^{+}}$, for every $j \in \set{1,\ldots, |C_{i}^{-}|}$ we add to $T$ the edges $\set{x_{i}^{'}, x_{ij}^{r-}}$, $\set{x_{ij}^{r-}, x_{ij}^{-}}$, $\set{x_{ij}^{-}, C_j}$ (note that for each clause $C_j$ we add only one such edge), and for every $j \in \set{1,\ldots, |C_{i}^{-}|-1}$ we add to $T$ the edges $\set{x_{ij}^{r-}, x_{i(j+1)}^{r-}}$.

Conversely, suppose now that $H$ contains an arborescence $T$ rooted at $r$ with cost at most $0$ and let us define a satisfying assignment of $\phi$. Since for every $i,j \in \set{4,5,6}$ with $i \neq j$ we have that $\cc(i,j)=1$, for every clause-vertex $C_j$ exactly one of its incident edges belongs to $T$. Since $\cc(i,8)=0$ and $\cc(i,7)=1$ for all $i\in\set{4,5,6}$, we have that if $\set{x_{ij}^{+},C_{j}}$ belongs to $T$, then $\set{x_{ij}^{+},x_{ij}^{r+}}$, $\set{x_{ij}^{r+},x_{i}^{r}}$, and all other edges with color 8 in the gadget corresponding to $C_{i}^{+}$ belong to $T$. Due to the same reasons, if $\set{x_{ij}^{-},C_{j}}$ belongs to $T$, then $\set{x_{ij}^{-},x_{ij}^{r-}}$, $\set{x_{ij}^{r-},x_{i}^{'}}$, and all edges with color 8 in the gadget corresponding to $C_{i}^{-}$ belong to $T$.
From the structure of $H$ and from the fact that $\cc(1,2)=1$, it follows that in order for the tree $T$ to span all vertices of $H$, for every $i \in \{1,\ldots,n\}$ either the three edges $\{x_i^{\ell}$, $x_i^{+}\}$, $\{x_i^{+},x_i^{-}\}$, $\{x_i^{-},x_i^{r}\}$ or the three edges $\{x_i^{\ell}$, $x_i^{-}\}$, $\{x_i^{-},x_i^{+}\}$, $\{x_i^{+},x_i^{r}\}$ belong to $T$. In the former case, we set variable $x_i$ to 1, and in the latter case we set variable $x_i$ to 0. Since for every $i \in \{4,5,6\}$ we have that $\cc(7,i)=1$, it follows that for every clause-vertex $C_j$, its incident edge that belongs to $T$ joins $C_j$ to a literal that is set to 1 by the constructed assignment. Hence, all clauses of $\phi$ are satisfied by this assignment, concluding the proof of the theorem.\end{proof}

\section{FPT algorithm for star tree-cutwidth}\label{sec:algorithms}
We have shown in Theorem~\ref{thm:hardness-treecutwidth}  that the $\prbcc$ problem is {\sc{W[1]}}-hard on multigraphs when parameterized by tree-cutwidth. In this section we present an $\fpt$ algorithm for a restricted version of this parameter. Besides being the first $\fpt$ algorithm for the problem (except for the one given in~\cite{FPT-by-tw-Delta}  parameterized by the treewidth and the maximum degree of the input graph), it provides us with the insight on the source of the problem's hardness.

Before stating the main result of this section, we introduce some preliminaries and notation used in this section. We start with some known facts about tree-cutwidth.

\begin{theorem}[Kim \emph{et al}.~\cite{KOPST15}]\label{thm:KimEtAl2Approx}
 Given a graph $G$ on $n$ vertices, a tree-cut decomposition of $G$ of width at most $2 \tcw(G)$ can be computed in time $2^{O({\scriptsize\tcw}(G)^2 \log {\scriptsize\tcw}(G))} \cdot n^2$.
\end{theorem}

A non-root node $t$ of $T$ is \emph{thin} if $\adh(t) \leq 2$ and \emph{bold} otherwise. A tree-cut decomposition $\tcd$ is \emph{nice} if for every thin node $t$ and every sibling $t'$ of $t$, $Y_t$ does not have neighbors in $Y_{t'}$. For a node $t$ of $T$ we use $B_t$ to denote the set of thin children $t'$  of $t$ such that $N(Y_{t'}) \subseteq X_t$, and we let $A_t$ contain every child of $t$ which is not in $B_t$.

\begin{theorem}[Ganian \emph{et al}.~\cite{Ganian0S15}]\label{thm:GanianAlgorithmNiceTCWD}
 A tree-cut decomposition of a graph $G$ can be transformed in time $O(n^3)$ to a nice tree-cut decomposition of $G$ without increasing its width or its number of nodes.
\end{theorem}

\begin{lemma}[Ganian \emph{et al}.~\cite{Ganian0S15}]\label{thm:GanianAtBounded}
For every node $t$ of a tree-cut decomposition of width $k$,
\begin{equation}\label{eq:At}
|A_t| \leq 2k+1.
\end{equation}
\end{lemma}

Given a graph $G$, a subset $U$ of its vertices, and a subset $F$ of its edges we denote by $G[U \cup F]$ the graph induced by these vertices and edges, that is obtained by adding to $G[U]$ the edges of $F$ and their endpoints. Formally, $G[U \cup F] \defined (U \cup V(F), F \cup E(G) \cap U \times U)$.

\begin{definition}
For a node $t$ of a given tree-cut decomposition of a graph $G$, and two distinct vertices $u,v$ of $X_t$, let $B_t^{u,v}$ be the set of children $t' \in B_t$ such that $N(Y_{t'})=\set{u,v}$. $\hat{G}_t$ is the multigraph obtained by augmenting $G[X_t]$ with a \emph{red} edge $uv$ for every $t' \in B_t^{u,v}$. A \emph{star tree-cut decomposition} is a tree-cut decomposition where $\hat{G}_t$ is the disjoint union of stars, possibly with parallel edges, for every node $t$ of the decomposition. The \emph{star tree-cutwidth} of $G$ is the smallest number $k \geq \tcw(G)$ such that every tree-cut decomposition of width at most $k$ is a star tree-cut decomposition whenever such a $k$ exists, and $n$ otherwise.
\end{definition}

We would like to emphasize that the above definition is somehow artificial, and we do not aim at claiming any practical application of it. The objective of the next theorem is to show that bounded tree-cutwidth is ``almost'' enough to provide an $\fpt$ algorithm for $\prbcc$, in the sense that if one imposes just a limited dependency among thin nodes (this is exactly what the `star' condition is about), then an $\fpt$ algorithm is possible. It may look clear to the reader that we were looking for an  $\fpt$ algorithm with parameter tree-cutwidth until we were able to prove Theorem~\ref{thm:hardness-treecutwidth}; nevertheless, we still think that the ideas behind the proof of Theorem~\ref{thm:FPT-tree-cut-width} are interesting.

\begin{theorem}\label{thm:FPT-tree-cut-width}
The $\prbcc$ problem is $\fpt$ parameterized by the star tree-cutwidth of the input graph.
\end{theorem}
\begin{proof} Let $(G,X,\chi,r,f)$ be an instance of $\prbcc$ with $|V(G)|=n$ and such that $\tcw(G) \leq k$. We first compute, using Theorem \ref{thm:KimEtAl2Approx}, a tree-cut decomposition $G$ of width at most $2k$, in time $2^{O(k^2 \log k)} \cdot n^2$. Then, using Theorem \ref{thm:GanianAlgorithmNiceTCWD} we transform it to a {\sl nice} tree-cut decomposition $\tcd$ of width at most $2k$, in time $O(n^3)$. In order to simplify the formulation, we add to $G$ a new vertex $r'$ and an edge $\set{r,r'}$. We also add a node $t_{r'}$ adjacent to the unique node $t_r$ of $T$ whose bag contains $r$. We root $T$ at this new node $t_{r'}$. Observe that the width of $\tcd$ is not affected by this transformation. We color the edge $\set{r,r'}$ with a new color $\hat{x}$ that does not incur any traversal cost, i.e., $f(x,\hat{x})=f(\hat{x},x)=0$ for every $x \in X$. Then, there is a one to one correspondence between the solutions of the original instance and the solutions of the modified instance. Moreover, the corresponding solutions have the same cost.

Following the general approach introduced in~\cite{Ganian0S15}, we aim to define a data structure $\dd_t$ at each node $t$ of $T$. The dynamic programming algorithm will perform a bottom-up traversal of $T$, and at each node $t$ of $T$ different than $t_{r'}$ it will compute the table $\dd_t$. Finally, we will extract an optimal solution from the table $\dd_{t_r}$. Since our goal is to find a spanning tree of $G$, we have to analyze how such a spanning tree is decomposed by the tree-cut decomposition $\tcd$.

%\begin{figure}[htbp]
%\begin{center}
%\commentfig{
%%\vspace{-3cm}
%%\hspace*{-2.5cm}
%\includegraphics[width=.9\textwidth,  angle =0]{figures/ForestDecomposition.pdf}%\vspace{-.5cm}
%}
%\caption{The decomposition of a forest $F_t$ associated with a node $t$ of a tree-cut decomposition} \label{fig:ForestDecomposition}
%\end{center}
%\end{figure}

A \emph{partial orientation} of a set of edges is an orientation of a subset of it. Given a partial orientation $\Phi$ of $\cut(t)$, we denote by $\Phi^+$ (resp., $\Phi^-$) the set of edges in $\cut(t)$ that are oriented toward (resp., outward) $Y_t$. In our algorithm, the set $\Phi$ will correspond to the set of edges in $\cut(t)$ that belong to the solution. We denote by $V(\Phi^+)$ (resp. $V(\Phi^-)$) the endpoints of the edges of $\Phi^+$ (resp.,$\Phi^-$) in $Y_t$.

Let $F_t$ be a rooted forest of $G[Y_t \cup \cut(t)]$ spanning all vertices in $Y_t$. The set of arcs $E(F_t) \cap \cut(t)$ is a partial orientation $\Phi$ of $\cut(t)$.  We say that $F_t$ \emph{induces} $\Phi$ on $\cut(t)$, or simply that $F_t$ induces $\Phi$ whenever $\cut(t)$ is clear from the context. We denote this by $F_t \Rightarrow \Phi$. $F_t$ is a \emph{good} forest if it has no sinks in $Y_t$. We note that a solution of the $\prbcc$ problem, being a spanning tree rooted at $r'$, induces a good forest on $Y_t$.

Let $t$ be a node of $T$ different than $t_{r'}$, $F_t$ a good forest that induces some partial orientation $\Phi_t$ on $\cut(t)$. Since $F_t$ does not have sinks in $Y_t$, every vertex of $Y_t$ has a directed path to a sink of $F_t$. For every vertex $v \in X_t$ the path from $v$ to its associated sink in $F_t$ contains at least one vertex in $V(\Phi_t^-)$. We now define a function $\varphi_t: V(\Phi_t^+) \to V(\Phi_t^-)$ that we will term as the \emph{partition} of $\Phi_t$ \emph{induced} by $F_t$. For every vertex $v \in V(\Phi_t^+)$, $\varphi_t(v)$ is the first vertex of $V(\Phi_t^-)$ on the path from $v$ to the corresponding sink of $F_t$. We also say that $F_t$ \emph{induces} the pair $(\Phi_t,\varphi_t)$. Clearly, a partition of $\Phi_t$ is equivalent to a forest on $V(\Phi_t^+) \cup V(\Phi_t^-)$ consisting of stars where the vertices of $V(\Phi_t^-)$ are the centers of the stars and the vertices of $V(\Phi_t^+)$ are the leaves. Note that whenever $v \in V(\Phi_t^-) \cap V(\Phi_t^-)$ we have $\varphi_t(v)=v$.
For a forest $F$ and $U \subseteq V(F)$, let $\rc_U(F)$ denote the cost of $F$ that is incurred on the vertices in $U$. Clearly, $\rc_U(F) \leq \rc(F)$.

At this point now ready to define the tables that we will use in our dynamic programming algorithm. At each node $t$ of $\tcd$, our table $\dd_t$ will store, for every partial orientation $\Phi_t$ of $\cut(t)$ and every partition $\varphi_t$ of $\Phi_t$, the minimum cost incurred on $Y_t$ of a forest inducing $(\Phi_t, \varphi_t)$, which we denote by $c(\Phi_t,\varphi_t)$. Note that $|\dd_t| \leq 3^{2k} \cdot (2k)^{2k} = 2^{O(k \log k)}$.
We note that $\cut(t_r)=\set{\set{r,r'}}$, thus every spanning tree rooted at $r'$ induces the partial orientation $\Phi_r$ that orients $\set{r,r'}$ from $r$ to $r'$, i.e., $\Phi_r^-=\set{\set{r,r'}}$. Since $\Phi_r^+=\emptyset$, then every spanning tree induces the empty partition. Therefore, the optimum of the instance is equal to the value $c(\Phi_r,\emptyset)$ in $\dd_{t_r}$.

We now proceed to describe how the values $c(\Phi,\varphi)$ can be recursively computed in a bottom-up way. Assume first that $t$ is a leaf of $T$, and fix a partial orientation $\Phi_t$ of $\cut(t)$. Let $F_t$ be a good forest of $t$ that induces $\Phi_t$. The forest $F_t$ can be constructed by an appropriate choice of a parent from $X_t$ to each vertex of $X_t \setminus V(\Phi_t^-)$. We  iterate over all such choices (at most $(2k)^{2k}$ in number), and for each choice we a) verify in $O(k)$ time, that the choice $F_t$ is indeed a forest, by comparing the number of connected components of $F_t$ to the number of connected components of $\Phi_t^-$, b) compute in $O(k)$ time the cost $\rc(F_t)$ of $F_t$ and finally set $c(\Phi_t,\varphi_t)=\min \set{c(\Phi_t,\varphi_t),\rc(F_t)}$ in $\dd_t$ where $\varphi_t$ is the partition induced by $F_t$. That is,
\[
c(\Phi_t,\varphi_t)= \min \set{\rc_{X_t}(F_t)~|~F_t \Rightarrow (\Phi_t, \varphi_t)}.
\]

The crucial part of the algorithm is how to perform the inductive step. That is, assuming that the tables for all the children of a node $t$ of $\tcd$ have been already computed, we have to show that the table at node $t$ can be computed in $\fpt$-time. Consult Fig. \ref{fig:ForestDecomposition} for the following discussion. Let $t$ be a non-leaf node of $T$, and $F_t$ be a forest of $G[Y_t \cup \cut(t)]$. $F_t$ induces
\begin{itemize}
\item[$\circ$] a partial orientation $\Phi_t$ of $\cut(t)$,
\item[$\circ$] a partial orientation $\Phi$ of $\delta(X_t)$,
\item[$\circ$] a partition $\varphi_t$ of $\Phi_t$,
\item[$\circ$] a forest $F$ of $G[X_t \cup \delta(X_t)]$, and
\item[$\circ$] for every child node $t'$ of $t$
\begin{itemize}
\item a forest $F_{t'}$ of $G[Y_{t'} \cup \cut(t')]$,
\item a partial orientation $\Phi_{t'}$ of $\cut(t')$,
\item a partition $\varphi_{t'}$ of $\Phi_{t'}$.
\end{itemize}
\end{itemize}
Therefore, we proceed as follows:
\begin{itemize}
\item[$\circ$] Guess a partial orientation $\Phi_t$ among the $3^{2k}=2^{O(k)}$ partial orientations of $\cut(t)$.
\item[$\circ$] Guess a partial orientation $\hat{\Phi}$ of $\cup_{t' \in A_t} \cut(t') \setminus \cut(t)$. Note that, since $\abs{A_k} \leq 4k+1$ by Equation (\ref{eq:At}), the number of possibilities is at most $3^{(4k+1)2k}=2^{O(k \log k)}$.
\item[$\circ$] $\Phi_t \cup \hat{\Phi}$ induces a partial orientation $\Phi$ of $\delta(X_t)$ and a partial orientation $\Phi_{t'}$ for every $t' \in A_t$.
\item[$\circ$] For every $t' \in A_t$ guess a partition $\varphi_{t'}$ of $\Phi_{t'}$. Note that the number of possible guesses is at most $((2k)^{2k})^{4k+1}=2^{O(k^2 \log k)}$.
\item[$\circ$] Guess a forest $F$ of $G[X_t \cup \delta(X_t)]$ that induces $\Phi$ by choosing a parent for each vertex of $X_t \setminus V(\Phi^-)$ and counting the number of connected components. The number of possible guesses is at most $(2k)^{2k}=2^{O(k \log k)}$.
\item[$\circ$] Verify that $\hat{F}=F \cup \left( \cup_{t' \in A_t} (\Phi_{t'} \cup \varphi_{t'}) \right)$ is a good forest.
\item[$\circ$] Let $\varphi_t$ be the partition that $\hat{F}$ induces on $\Phi_t$.
\end{itemize}

\begin{figure}[t]
\begin{center}
\commentfig{
\vspace{-.3cm}
%\hspace*{-2.5cm}
\includegraphics[width=.99\textwidth,  angle =0]{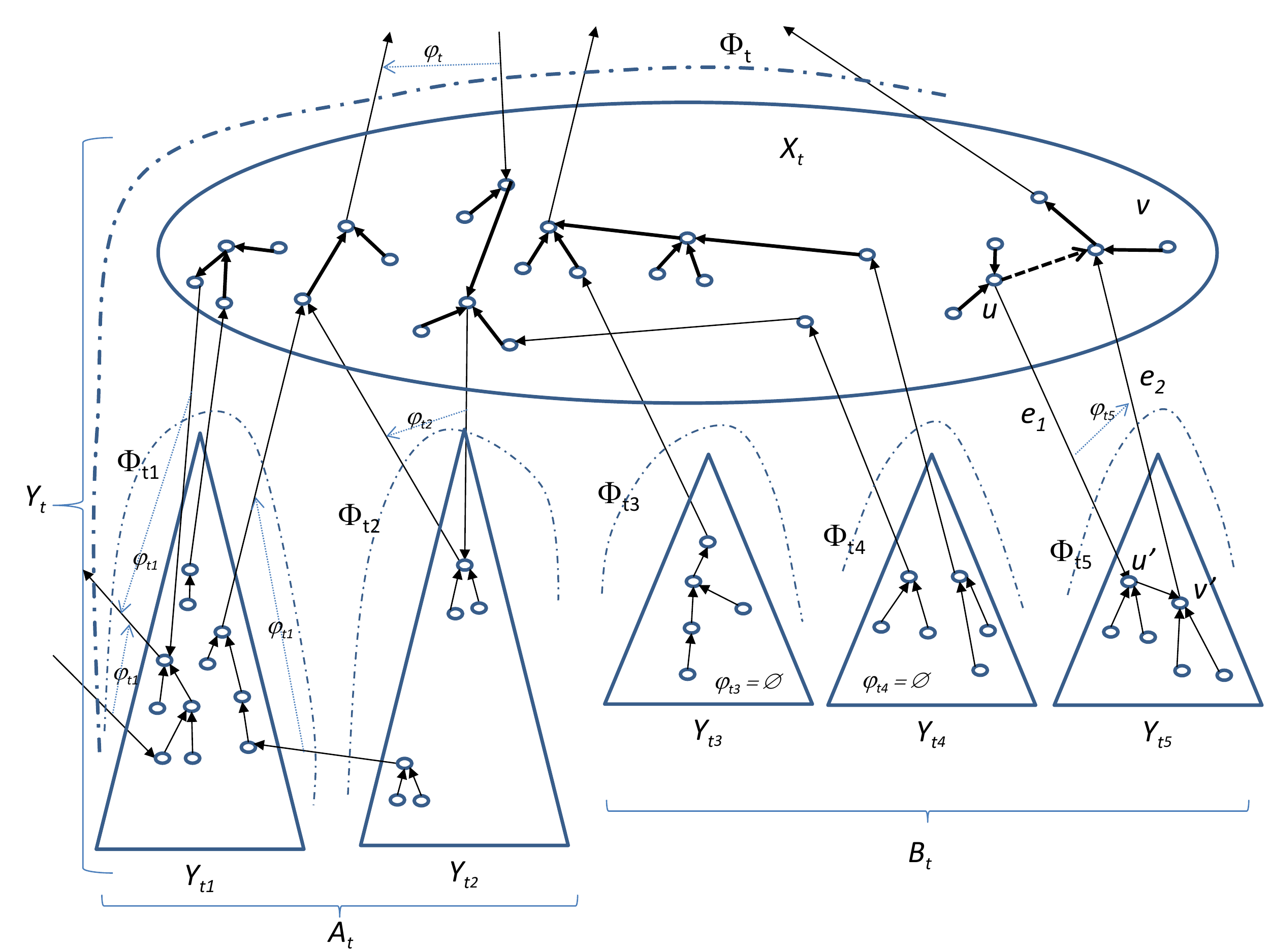}\vspace{1.cm}
}\vspace{-.5cm}
\caption{The decomposition of a forest $F_t$ associated with a node $t$ of a tree-cut decomposition in the proof of Theorem~\ref{thm:FPT-tree-cut-width}.} \label{fig:ForestDecomposition}
\end{center}
\end{figure}

First, we assume for simplicity that $B_t=\emptyset$. It is important to note that $F_t$ induces both $F$ and $(\Phi_t, \varphi_t)$ if and only if $F_t = F \cup \left( \cup_{t' \in A_t} F_{t'} \right)$ where each forest $F_{t'}$ induces $(\Phi_{t'},\phi_{t'})$ on $\cut(t')$. Moreover, $\rc_{Y_t}(F_t)=\rc_{X_t}(F)+\sum_{t' \in A_t} \rc_{Y_{t'}}(F_{t'})$. Therefore, the minimum cost forest $F_t$ that can be obtained by this set of guesses is
\begin{equation}\label{eqn:MinCostForest}
\hat{c}=\rc_{X_t}(F) + \sum_{t' \in A_t} c(\Phi_{t'},\varphi_{t'}).
\end{equation}
Finally, we set $c(\Phi_t,\varphi_t)=\min \set{c(\Phi_t,\varphi_t),\hat{c}}$ in the table $\dd_t$.

To this end, our algorithm is to iterate over all guesses above, for each valid guess to compute its cost $\hat{c}$ and to store for each pair $(\Phi_t, \varphi_t)$ the smallest value $\hat{c}$ associated with this pair. We note that the number of guesses is at most $2^{O(k^2 \log k)}$, and that the computations involved in each guess can be performed in polynomial time.

It remains to deal with the children of $t$ in the set $B_t$, which can be arbitrarily many (that is, not necessarily bounded by a function of $k$). Recall that for every such child $t' \in B_t$, it holds that $\adh(t') \leq 2$ and that $N(Y_{t'}) \subseteq X_t$. Since we can assume that $G$ is connected, we have that $\adh(t') \geq 1$ for every child $t' \in B_t$. Let $t' \in B_t$ such that $\adh(t') = 1$, and let $\cut(t') = \set{e_{t'}}$. Clearly, the edge $e_{t'}$ belongs to any arborescence spanning $G$, in particular to $F_t$. In other words the only partial orientation to be considered for $\cut(t')$ is $\Phi_{t'}^-= \set{e_{t'}}$ and the only possible partition is $\emptyset$. It is sufficient to add the term $c(\Phi_{t'}^-,\emptyset)$ to the right hand side of Equation (\ref{eqn:MinCostForest}).

We now consider $t' \in B_t$ with $\adh(t') = 2$. Assume that $t'$ is such that both edges in $\cut(t')$, say $e_1$ and $e_2$, are incident to the same vertex $v$ in $X_t$. Clearly, none of $e_1$ and $e_2$ can be oriented away from $X_t$, and therefore there are only three potential partial orientations of $\cut(t')$ that can possibly be induced by $F_t$ are $\Phi_{t',1}^- = \set{e_1}$, $\Phi_{t',2}^- = \set{e_2}$, or $\Phi_{t',3}^- = \set{e_1, e_2}$, and $\Phi_{t',1}^+= \Phi_{t',2}^+ = \Phi_{t',3}^+ =\emptyset$. Thus $\varphi_{t',1}=\varphi_{t',2}=\varphi_{t',3}=\emptyset$. Let $e_v$ be the edge leading $v$ to its parent in $\hat{F}$. Then, for $i \in [3]$ the $i$-th possibility contributes the value $c(\Phi_{t',i},\emptyset) + \sum_{e_i \in \Phi_{t',i}^-} f(e_v,e_i)$. Therefore, we add to the right hand side of Equation (\ref{eqn:MinCostForest}) the term
\[
\min \{c(\Phi_{t',i},\emptyset) + \sum_{e_i \in \Phi_{t',i}^-} f(e_v,e_i) |~1 \leq i \leq 3 \}.
\]

To this end our algorithm remains intact, except the computation of $\hat{c}$ for each guess that now contains additional terms. We remain with the case that the two edges of $\cut(t')$ are incident to two distinct vertices $u,v$ of $X_t$. Recall that in this case $t' \in B_t^{u,v}$ (the node $t'_5$ in Fig.~\ref{fig:ForestDecomposition} is an example of such a node).

Let $\cut(t')=\set{e_1,e_2}$ where $e_1=\set{u,u'}$ and $e_2=\set{v,v'}$ (note that possibly $u'=v'$). In this case there are five potential partial orientations of $\cut(t')$ that $F_t$ can possibly induce, namely the partial orientations $\Phi_{t',1},\Phi_{t',2},\Phi_{t',3}$ as in the previous case, and two partial orientations $\Phi_{t',uv}$ and $\Phi_{t',vu}$ that orient one edge towards $Y_t$ and the other towards $X_t$. For each one of the last two cases there is exactly one possible partition: $\varphi_{t',uv}$ and $\varphi_{t',vu}$, respectively. We consider two cases.
\begin{itemize}
\item[$\circ$] $F_{t'}$ induces $(\Phi_{t',i}, \varphi_{t',i})$ for some $i \in [3]$: In this case we can choose the best value for $i$ as we did before.

\item[$\circ$] $F_{t'}$ induces $(\Phi_{t',uv},\varphi_{t',uv})$ where $\Phi_{t',uv}$ orients $e_1$ from $u$ to $u'$ and $e_2$ from $v'$ to $v$: We will modify $F_t$ to obtain another "equivalent" forest which does not span $Y_{t'}$ but contains edges that not present in $G$. We remove $F_{t'}$ from $F_t$ and replace it by a \emph{simulating} arc $\hat{e}_{t',uv}$ from $u$ to $v$. Furthermore, we assign to this arc a weight $w(\hat{e}_{t',uv})=c_{t'}(\Phi_{t',uv},\varphi_{t',uv})$ that corresponds to the cost of traversing the entire tree $F_{t'}$ (starting from its root $u$ and ending at the leaves including $v$), i.e., $c(\Phi_{t',uv}, \varphi_{t',uv})$. In order to simulate the two traversal costs at the vertices $u$ and $v$, we assign to $\hat{e}_{t',uv}$ a unique color $x_{t',uv}$ such that the cost of entering (resp., leaving) this edge is equal to the cost of entering $e_1$ (resp., $e_2$). That is, $f(x,x_{t',uv})=f(x,\chi(e_1))$ and $f(x_{t',uv},x)=f(\chi(e_2),x)$ for every color $x \in X$. Let $\hat{F}_t$ be the forest obtained by repeating this transformation for every pair $u,v$ and every $F_{t'}$ that induces one of $(\Phi_{t',uv},\varphi_{t',uv})$, $(\Phi_{t',vu},\varphi_{t',vu})$. By the construction, we have that $\rc_{X_t}(F_t)=\rc_{X_t}(\hat{F}_t) + w(\hat{F}_t)$. Note that the simulating arcs correspond to the red edges of $\hat{G}_t$.
\end{itemize}

Therefore, it would be sufficient to modify the way $F$ is guessed so that it allows $F$ to contain red edges of $\hat{G}_t$. However, since the number of these edges is not bounded by a function of $k$, this would not imply an $\fpt$.

To cure this problem, we define $\hat{\hat{G}}_t$ as the multi-graph obtained from $\hat{G}_t$ by replacing every set of parallel red edges by one red edge. We also allow $F$ to contain a red edge of $\hat{\hat{G}}_t$. Since the number of edges of $\hat{\hat{G}}_t$ is bounded by a function of $k$, the number of choices for $F$ remains a function of $k$. In the sequel, our goal is to find the best forest $F_t$ that induces $F$ and all the guessed values. In other words we want to find a) a red edge of $\hat{G}_t$ for each red edge of $F$, and b) for every other red edge of $\hat{G}_t$ to decide upon one of the possible partial orientations.

In the sequel we show how to make these decisions (namely a) and b) above) using a dynamic programming algorithm that performs a bottom-up traversal of $F$. We first make the following simplifying assumption: every red edge of $\hat{\hat{G}}$ is in $F$. For $v \in X_t$, let $F_{t,v}$ be the subtree of $F_t$ rooted at $v$. Let $e=(u,v)$ be an arc of $F$, and $\hat{e}$ be any edge of $\hat{G}$ between $u$ and $v$. We denote by $F_{t,\hat{e}}$ the tree consisting of $F_{t,v}$ and the arc $\hat{e}$. We discard costs incurred by traversals in vertices in subtrees $Y_{t'}$ since they are fixed by the current set of guesses. We compute in a bottom-up fashion the values $OPT_F(\hat{e})$ that denote the minimum cost of the subtree $F_{t,\hat{e}}$ among all trees $F_t$ that agree with the current set of guesses. Note that for a node $t' \in B_t^{u,v}$, once the inbound edges of $u$ and $v$ are fixed, we can decide on the best partial orientation $\Phi_{t'}$ by comparing the three possible values.

If $v$ is a leaf of $F$ then $F_{t,\hat{e}}$ contains at most one type a) red edge, namely $\hat{e}$, and no other red edges, by our assumption. Therefore, $OPT_F(\hat{e})$ can be computed by summing up the traversal costs between $\hat{e}$ and all the edges of $\hat{\Phi}$ oriented from $v$ outside. If $v$ is not a leaf of $F$, let $e_1, \ldots, e_\ell$ be the arcs from $v$ to its children. For each such arc $e_j$ we have to choose exactly one arc $\hat{e_j}$ of $\hat{G}$. The crucial point is that these choices can be made independently of each other. For every possible choice for $\hat{e_j}$ we compute the cost incurred by traversing $v$ from $\hat{e}$ to $\hat{e_j}$ and for each arc $\hat{e_{j'}} \neq \hat{e_j}$ we add the minimum among the three possible costs. Note that these costs can be computed since the two edges incident to $\hat{e_{j'}}$ are known (namely $\hat{e}$ and $\hat{e_j}$) at this point. We choose the arc $\hat{e_j}$ that leads to the smallest cost, and we repeat this for every edge $e_j$ and sum up the costs.

Finally, it remains to relax the simplifying assumption, i.e., to handle the edges of $\hat{\hat{G}}$ that are not in $F$. Let $\hat{e}=u$ be an edge of $\hat{G}$ that corresponds to such an edge. There are only three possible ways to traverse the subtree corresponding to $\hat{e}$, in order to choose the best one we have to know the edges $e_u, e_v$ of $F_t$ leading $u$ and $v$ to their parents, respectively. Since the red edges of $\hat{G}$ constitute a union of stars, at least one of $e_u, e_v$ is not red, therefore its color is known. Then, the cost of $\hat{e}$ depends only on the choice of one (red) edge. We can modify the above dynamic programming algorithm so that the cost of each such edge is associated with the red edge in $F$ incident to it. If no such edge exists, then the cost of $\hat{e}$ is constant for the current choice of $F$.
\end{proof}

\newpage
\section{Conclusions and further research}\label{sec:conclusion}
In this article we proved several hardness results for the $\prbcc$ problem. In particular, we proved that the problem is {\sc{W[1]}}-hard parameterized by treewidth on general graphs, and that it is {\sc{NP}}-hard on planar graphs, but we do not know whether it is {\sc{W[1]}}-hard parameterized by treewidth (or treedepth) on planar graphs.

On the other hand, we provided an {\sc{FPT}} algorithm for a restricted version of tree-cutwidth, and we proved that the problem is {\sc{W[1]}}-hard on multigraphs parameterized by tree-cutwidth. While we were not able to prove this {\sc{W[1]}}-hardness result on graphs without multiple edges, we believe that it is indeed the case. It would be natural to consider other structural parameters such as the size of a vertex cover or a feedback vertex set.

Finally, it would be interesting to try to generalize our techniques to prove hardness results or to provide efficient algorithms for other reload cost problems that have been studied in the literature~\cite{WiSt01,Ga08,CGM13,GGM14}.

\bigskip

{\small\noindent\textbf{Acknowledgement}. We would like to thank the anonymous referees of the conference version of this article for helpful comments that improved the presentation of the manuscript.}

\bibliographystyle{abbrv}	
\bibliography{reload-FPT}

\end{document}